\documentclass[lettersize,journal]{IEEEtran}
\usepackage{booktabs}
\usepackage{amsmath,amsfonts}
\usepackage{array}
\usepackage[caption=false,font=scriptsize,labelfont=rm,textfont=rm]{subfig}
\usepackage{textcomp}
\usepackage{stfloats}
\usepackage{url}
\usepackage{verbatim}
\usepackage{graphicx}
\usepackage{cite}
\usepackage{CJKutf8}
\usepackage{amsthm,amsmath,amssymb} 
\usepackage{algpseudocode}   
\usepackage{mathrsfs}
\usepackage{multicol}
\usepackage{svg}
\usepackage{graphicx}
\usepackage[ruled,linesnumbered]{algorithm2e}
\usepackage{algpseudocode}
\usepackage[colorlinks,linkcolor=blue,anchorcolor=blue,citecolor=blue]{hyperref}

\begin{document}

\title{Stacked Intelligent Metasurface for End-to-End OFDM System}

\author{Yida Zhang, Qiuyan Liu, Hongtao Luo, Yuqi Xia, Qiang Wang,~\IEEEmembership{Member,~IEEE}, Fuchang Li,\\ Xiaofeng Tao,~\IEEEmembership{Senior Member,~IEEE}, and Yuanwei Liu,~\IEEEmembership{Fellow,~IEEE}
        % <-this % stops a space
\thanks{This paper was partially funded by the National Key R \& D
Program of China (2020YFB1806602), BUPT-China Unicom Joint Innovation Center and Fundamental Research Funds for the Central Universities (2242022k60006) \textit{(Corresponding author: Qiang Wang, Qiuyan Liu.)}.}
\thanks{Yida Zhang, Hongtao Luo, Yuqi Xia, Qiang Wang, Xiaofeng Tao are with the National Engineering Research Center for Mobile Network Technologies, Beijing University of Posts and Telecommunications, Beijing 100876, China (e-mail: \{zhangyida02, mashirokaze1971, xiayuqi, wangq, taoxf\}@bupt.edu.cn).}
\thanks{Qiuyan Liu, Fuchang Li are with the China United Network Communications Corporation Research Institute, Beijing 100037, China (e-mail: \{liuqy95, lifc50\}@chinaunicom.cn).}
\thanks{Yuanwei Liu is with the Department of Electrical and Electronic Engineering, The University of Hong Kong, Hong Kong (email: yuanwei@hku.hk).}
\thanks{Simulation code and picture: }
}
%\url{https://github.com/ZhangYida02/Stacked_Intelligent_Metasurfaces_for_End-to-End_OFDM_System}.
\maketitle

\begin{abstract}
%堆叠智能超表面（SIM）和双极化SIM(DPSIM)使能的波域信号处理有望成为卸载基带数字处理任务，高效简化收发机设计的一种有潜力的研究方向。然而，现有的架构仅限于使用SIM (DPSIM)来实现单一的通信任务，如预编码与合并。为进一步提高SIM (DPSIM)辅助系统的整体性能，实现发送比特流到接收比特流全链路联合优化。我们提出了SIM (DPSIM)辅助的端到端（E2E)正交频分复用（OFDM）系统，其中信道编码、调制、预编码、合并、解调、信道解码等传统通信任务在电磁（EM）前向传播中同步进行。进一步的，我们通过将真实超表面抽象为网络隐藏层的思想，提出了电磁神经网络（EMNN），以实现对E2E OFDM通信系统的驱动控制。此外，我们将迁移学习引入模型训练，设计了EMNN的一个训练和部署框架。仿真结果表明，SIM辅助E2E OFDM系统和DPSIM辅助E2E OFDM系统能够在复杂信道条件下实现鲁棒的比特流传输。我们的研究充分展示了EMNN和SIM辅助E2E OFDM系统在新一代收发机设计中的应用潜力。
Stacked intelligent metasurface (SIM) and dual-polarized SIM (DPSIM) enabled wave-domain signal processing have emerged as promising research directions for offloading baseband digital processing tasks and efficiently simplifying transceiver design. However, existing architectures are limited to employing SIM (DPSIM) for a single communication function, such as precoding or combining. To further enhance the overall performance of SIM (DPSIM)-assisted systems and achieve end-to-end (E2E) joint optimization from the transmitted bitstream to the received bitstream, we propose an SIM (DPSIM)-assisted E2E orthogonal frequency division multiplexing (OFDM) system, in which traditional communication tasks such as channel coding, modulation, precoding, combining, demodulation, and channel decoding are performed synchronously within the electromagnetic (EM) forward propagation. Furthermore, inspired by the idea of abstracting real metasurfaces as hidden layers of a neural network, we propose the EM neural network (EMNN) to enable the control of the E2E OFDM communication system. In addition, transfer learning is introduced into the model training, and a training and deployment framework for the EMNN is designed. Simulation results demonstrate that both SIM-assisted E2E OFDM systems and DPSIM-assisted E2E OFDM systems can achieve robust bitstream transmission under complex channel conditions. Our study highlights the application potential of EMNN and SIM (DPSIM)-assisted E2E OFDM systems in the design of next-generation transceivers.
\end{abstract}

\begin{IEEEkeywords}
Stacked intelligent metasurfaces (SIM), Dual-polarized stacked intelligent metasurfaces (DPSIM), Orthogonal frequency-division multiplexing (OFDM), electromagnetic neural network (EMNN), End-to-end (E2E).
\end{IEEEkeywords}

\section{Introduction.} 
\subsection{Research background.}
%先进的收发器设计已成为实现第六代（6G）无线网络中极高可靠性、超高速数据传输以及广泛移动连接的关键技术支撑之一~\cite{You2021Towards6G},~\cite{Cui2025OverviewAI6G}。在过去几十年里，多输入多输出（MIMO）天线架构有了显著的发展以实现更大的空间分集和复用增益~\cite{7959183}。传统的全数字天线阵列通过基带数字预编码和组合来减轻天线间干扰（IAI），从而确保数据流的独立传输~\cite{4549739}。然而，这种方法需要大量的射频（RF）链路，导致能耗和硬件成本过高，这给资源受限的终端带来了挑战~\cite{5783993}。对此，一种经典的手段是采用数字模拟混合架构，其尝试通过移相网络来减少RF链路数量，试图在性能和成本之间取得平衡~\cite{8030501},~\cite{7880698}。
Advanced transceiver design has emerged as a critical enabler for realizing ultra-high reliability, ultra-high-speed data transmission, and ubiquitous mobile connectivity in sixth-generation (6G) wireless networks~\cite{You2021Towards6G,Cui2025OverviewAI6G}. Over the past decades, multiple-input multiple-output (MIMO) antenna architectures have evolved substantially to provide higher spatial diversity and multiplexing gains~\cite{7959183}. In conventional fully-digital antenna arrays, baseband digital precoding and combining are employed to mitigate inter-antenna interference (IAI), thereby ensuring the independent transmission of multiple data streams~\cite{4549739}. However, such an approach requires a large number of radio frequency (RF) chains, resulting in excessive power consumption and hardware cost, which becomes prohibitive for resource-constrained terminals~\cite{5783993}. To alleviate this issue, a well-established solution is to adopt a hybrid digital–analog architecture, wherein the number of RF chains is reduced via a phase-shifting network, thereby aiming to achieve a favorable trade-off between performance and cost~\cite{8030501,7880698}.

%最近，可重构智能天线表面（RIS）作为一种关键技术应运而生，它能够在波域直接操控电磁（EM）波，从而有可能将传统上由射频链路处理的大量信号处理任务卸载出去~\cite{9424177}。在此基础上，一种将 RIS 直接集成在收发器中的新型天线架构被提出~\cite{9133266},~\cite{9779086}。通过精确控制 RIS 单元的电磁响应，收发器能够在数字域和波域实现联合信号处理。这种集成架构被称为全息多输入多输出（HMIMO）~\cite{10819473}。HMIMO 有效地减少了所需的射频链路数量，并通常采用低分辨率数字-模拟转换器（DAC）和模拟-数字转换器（ADC），从而显著降低了硬件成本并提高了能效~\cite{10232975}。
Recently, reconfigurable intelligent surfaces (RISs) have emerged as a pivotal technology capable of directly manipulating electromagnetic (EM) waves in the wave domain, thereby potentially offloading a substantial portion of the signal processing tasks traditionally handled by RF chains~\cite{9424177}. Building upon this concept, a novel antenna architecture has been proposed in which the RIS is directly integrated into the transceiver~\cite{9133266,9779086}. By precisely controlling the EM responses of the RIS elements, the transceiver can perform joint signal processing in both the digital and wave domains. This integrated architecture, referred to as holographic MIMO (HMIMO)~\cite{10819473}, effectively reduces the number of required RF chains and typically employs low-resolution digital-to-analog converters (DACs) and analog-to-digital converters (ADCs), thereby significantly lowering hardware costs and improving energy efficiency~\cite{10232975}.

%然而，受限于RIS的单层结构，其对EM波的操纵灵活性终究有限~\cite{10534211}。为进一步提升波域信号处理能力，~\cite{10158690} 和~\cite{11014597} 提出了堆叠智能超表面（SIM）辅助的HMIMO系统，与传统的大规模MIMO系统和RIS辅助的HMIMO系统相比，显著提升了系统性能。具体来说，SIM是一种三维超表面器件，其利用级联的多层超表面来实现EM波域的模拟计算~\cite{liu2022programmable}。在实践中，SIM通常被集成在RF前端从而调节发射或接收信号的振幅和相位，从而替代传统架构中的预编码（合并）模块，最终达到简化收发器设计的效果~\cite{10515204}。此外，~\cite{11091527}提出可以将双极化RIS进行堆叠，从而实现在两个正交极化方向上独立处理信号。这一架构被称为双极化SIM（DPSIM），其进一步提升了有限集成空间前提下的信号处理能力。
However, due to the inherent single-layer structure of RISs, their flexibility in manipulating EM waves remains limited~\cite{10534211}. To further enhance wave-domain signal processing capabilities,~\cite{10158690,11014597} proposed HMIMO systems assisted by stacked intelligent metasurfaces (SIMs), which significantly outperform conventional massive MIMO systems and RIS-assisted HMIMO systems. Specifically, a SIM is a three-dimensional metasurface device that leverages cascaded multilayer metasurfaces to perform analog computing in the EM wave domain~\cite{liu2022programmable}. In practice, SIMs are typically integrated into the RF front end to adjust the amplitude and phase of the transmitted or received signals, thereby replacing the precoding (or combining) modules in traditional architectures and ultimately simplifying transceiver design~\cite{10515204}. Furthermore, ~\cite{11091527} proposed stacking dual-polarized RISs to enable independent signal processing along two orthogonal polarization directions. This architecture, referred to as the dual-polarized SIM (DPSIM), further enhances signal processing capability under the constraint of limited integration space.

%在上述硬件架构持续演进的同时，基于端到端（E2E）学习的信号处理范式正逐渐成为提升下一代无线系统全链路性能的有力工具~\cite{8214233}。不同于传统的分模块优化方法，E2E学习通过在收发两端部署神经网络来对物理层的编码、调制、波束赋形、检测等过程进行联合建模与优化，从而在给定硬件约束下逼近全局最优性能\cite{8054694}。对于RIS或SIM辅助的HMIMO收发器，E2E学习能够自适应地在数字域与波域共同进行参数配置~\cite{9745781},~\cite{teng2025basebandfreeendtoendcommunicationbased}。这不仅有助于充分挖掘超表面的潜能，还能在存在硬件非理想条件时（如相位噪声、有限量化动态范围）保持较强的鲁棒性，因此被视为实现智能化6G收发器设计的重要方向\cite{8985539},\cite{9360873}。  
Along with the continuous evolution of hardware architectures, the end-to-end (E2E) learning-based signal processing paradigm is gradually becoming a powerful tool for enhancing the full-chain performance of next-generation wireless systems~\cite{8214233}. Different from the traditional modular optimization approach, E2E learning jointly models and optimizes physical layer processes such as coding, modulation, beamforming, and detection by deploying neural networks at both the transmitter and receiver, thereby approaching globally optimal performance under given hardware constraints~\cite{8054694}. For RIS- or SIM-assisted HMIMO transceivers, E2E learning can adaptively configure parameters in both the digital and wave domains~\cite{9745781},~\cite{teng2025basebandfreeendtoendcommunicationbased}. This not only helps fully exploit the potential of metasurfaces but also maintains strong robustness under hardware impairments (e.g., phase noise, limited quantization dynamic range). Therefore, it is regarded as an important direction for achieving intelligent 6G transceiver design~\cite{8985539},\cite{9360873}.

\subsection{Related work.} 
\textbf{SIM:} 
%继~\cite{10158690} 和~\cite{11014597}提出可以使用SIM来在波域完成信号处理任务以来，一些关于SIM进一步的研究陆续出现。~\cite{10922857} 探究了 SIM 在多用户波束成形中的应用，实现了比传统 MIMO 系统更高的频谱效率（SE）和更低的能耗~\cite{li2025stackedintelligentmetasurfacesenhancedmimo} 研究了 SIM 在宽带 HMIMO 系统中的潜力，在频率选择性衰落条件下实现了无干扰多载波传输。~\cite{11032184} 提出了一种用于实现速率最大化和最小化子信道干扰（ISI）的混合优化方法，证实了增加超表面层的数量有助于提升整个系统性能。~\cite{11036161} 利用 Z 参数建立了 SIM 的多端口网络模型，以克服传统简化方法导致的单向传播和忽略互耦等问题。此外，SIM 已被应用于诸如集成传感与通信（ISAC）~\cite{10803090} 和语义通信~\cite{10755151} 等领域。
Since the works in~\cite{10158690,11014597} first proposed utilizing SIMs to accomplish signal processing tasks in the wave domain, a series of follow-up studies on SIMs have emerged. The authors of~\cite{10922857} investigated the application of SIMs in multi-user beamforming and achieved higher spectral efficiency and lower energy consumption compared with conventional MIMO systems. The study in~\cite{li2025stackedintelligentmetasurfacesenhancedmimo} explored the potential of SIMs in wideband HMIMO systems, enabling interference-free multicarrier transmission under frequency-selective fading. In~\cite{11032184}, a hybrid optimization method was proposed to achieve rate maximization and minimize inter-symbol interference, demonstrating that increasing the number of metasurface layers can improve the overall system performance. The work in~\cite{11036161} developed a multiport network model for SIMs based on Z-parameters to address issues such as unidirectional propagation and neglected mutual coupling in conventional simplified approaches. Furthermore, SIMs have been applied to various domains, including integrated sensing and communication (ISAC)~\cite{10803090} and semantic communications~\cite{10755151}. 

\textbf{DPSIM:} 
%关于DPSIM的相关研究也正如火如荼。作为SIM的改进版，DPSIM由DPRIS堆叠而成，因能在两个正交极化方向上独立处理信号，从而拥有相较SIM更强的信号处理能力。~\cite{11091527}首次提出了DPSIM辅助的HMIMO系统，评估了在相同集成空间下HMIMO系统的性能提升。~\cite{10847939} 结合速率分割提出了一种的DPSIM收发器架构（RS-DPSIM），其在两个极化方向上同时分别传输公共信息和私有信息，最后仿真展示了该架构在抗干扰通信中的性能优势。
As an improved version of SIM, DPSIM is constructed by stacking DPRISs and can independently process signals along two orthogonal polarization directions, thereby achieving stronger signal-processing capability than SIM. ~\cite{11091527} first proposed a DPSIM-assisted HMIMO system and evaluated the performance gains achievable within the same integration footprint. Furthermore,~\cite{10847939} proposes a DPSIM-assisted transceiver architecture combined with rate-splitting (RS), which simultaneously transmits common and private information along two polarization directions. Simulations ultimately demonstrate the performance advantages of this architecture in anti-jamming communications.

\textbf{E2E learning of OFDM system:} 
% 在过去的几年里，大量研究工作聚焦于端到端（E2E）学习在不同通信系统中的性能探索。~\cite{8445920} 首次将自动编码器（AE）引入正交频分复用（OFDM）系统。然而，其假设每个子载波独立映射，该设计限制了接收端对信道时频相关性的利用。针对这一问题，~\cite{8052521}优化了基于全连接层的神经网络接收机，以同时联合处理多个子载波，这一方案显著提升了系统性能。~\cite{8663458}探索了量化比特对OFDM系统E2E通信性能的影响，并验证了基于学习方法在各种复杂信道条件下的鲁棒性。~\cite{9508784} 则针对频率与时间选择性衰落信道下的OFDM波形E2E通信展开研究，展示了基于导频和星座点联合学习的性能增益。此外，一些新型网络结构（如生成对抗网络（GAN）~\cite{8985539} 和深度全卷积神经网络（DeepRx）~\cite{9345504}）也被应用在E2E通信中，以简化收发网络设计。
In recent years, extensive research has focused on exploring the performance of E2E learning in various communication systems. ~\cite{8445920} was the first to introduce autoencoders (AEs) into orthogonal frequency-division multiplexing (OFDM) systems. However, it assumes independent mapping on each subcarrier, which limits the receiver’s ability to exploit the time–frequency correlation of the channel. To address this issue, ~\cite{8052521} optimized a neural-network receiver to jointly process multiple subcarriers, significantly improving system performance. ~\cite{8663458} investigated the impact of quantized bits on the E2E communication performance of OFDM systems and verified the robustness of learning-based approaches under various complex channel conditions. ~\cite{9508784} studied E2E communication with OFDM waveforms over frequency- and time-selective fading channels, demonstrating performance gains from jointly learning pilots and constellation points. In addition, several novel network architectures—such as generative adversarial networks (GANs) ~\cite{8985539} and deep fully convolutional neural networks (DeepRx) ~\cite{9345504} have also been applied to E2E communications to simplify transceiver design.

\subsection{Motivation and contribution.}
%SIM和DPSIM能够在射频前端对信号进行波域计算，从而实现基带处理任务的卸载。凭借卓越的能效和几乎忽略不计的时延，波域信号处理有望成为简化收发机设计的重要研究方向~\cite{11014597}。然而，目前的研究仅限于使用SIM和DPSIM来取代通信链路里的单一模块（一般为预编码或合并），而基带仍然负责调制、检测等其他信号处理任务，如此分模块的优化难以达到整个系统的最佳性能~\cite{teng2025basebandfreeendtoendcommunicationbased}。此外，由于整个收发架构的变革，如何对SIM (DPSIM)辅助HMIMO系统和传统massive MIMO系统进行公平性能比较成为尚待解决的问题~\cite{10158690},~\cite{li2025stackedintelligentmetasurfacesenhancedmimo},~\cite{11091527}。为应对如上挑战，受RIS辅助E2E系统启发~\cite{9745781}，我们提出了SIM (DPSIM)辅助的E2E OFDM 系统，并设计了一个完整的的训练和部署框架。The main contributions of this paper are as follows:
SIM and DPSIM are capable of performing wave-domain computations at the RF frontend, thereby enabling the offloading of baseband processing tasks. Owing to their outstanding energy efficiency and negligible latency, wave-domain signal processing is envisioned as a promising research direction to simplify transceiver design~\cite{11014597}. However, existing studies are limited to replacing a single module in the communication link (typically precoding or combining) with SIM or DPSIM, while the baseband still undertakes other signal processing tasks such as modulation and detection. Such modular optimization makes it difficult to achieve the optimal performance of the entire system~\cite{teng2025basebandfreeendtoendcommunicationbased}. In addition, due to the architectural transformation of the transceiver, how to fairly compare the performance of SIM (DPSIM)-assisted HMIMO systems with that of conventional massive MIMO systems remains an open problem~\cite{10158690},~\cite{li2025stackedintelligentmetasurfacesenhancedmimo},~\cite{11091527}. To address the above challenges, inspired by the RIS-assisted E2E systems~\cite{9745781}, we propose an SIM (DPSIM)-assisted E2E OFDM system and develop a complete training and deployment framework. The main contributions of this paper are summarized as follows:

\begin{enumerate}[]
    \item %我们首次建立了SIM辅助E2E OFDM系统和DPSIM辅助E2E OFDM系统的数学模型。与以往的SIM (DPSIM)研究不同，该系统旨在实现通信链路从发射比特流到接收比特流的联合优化，其通过部署DNN和集成SIM (DPSIM)器件，实现了信号处理任务从基带数字域到波域的完全卸载。
    We are the first to establish mathematical models for SIM-assisted E2E OFDM systems and DPSIM-assisted E2E OFDM systems. Different from existing studies on SIM (DPSIM), the proposed system aims to achieve joint optimization of the communication link from the transmitted bit stream to the received bit stream. By deploying deep neural networks (DNNs) and integrating SIM (DPSIM) devices, the systems enable complete offloading of signal processing tasks from the baseband digital domains to the wave domains.  
    \item %为整体优化E2E系统性能，我们通过将真实超表面抽象为网络隐藏层，进一步提出了电磁神经网络（EMNN）。EMNN将隐藏层参数与EM（DPEM）单元相联系，从而实现了模型驱动控制的效果。
    To achieve holistic optimization of the E2E system performance, we further propose the EM neural network (EMNN) by abstracting the physical metasurface as hidden layers in the network. The EMNN associates the hidden-layer parameters with the EM (DPEM) units, thereby enabling model-driven control.
    \item %为了提高EMNN的训练速度，我们将迁移学习引入模型训练，通过统计CSI来训练基础网络，然后基于瞬时CSI来对网络微调。最后，我们提出了一个EMNN的训练和部署框架。
    To accelerate the training of EMNN, we introduce transfer learning into the model training process, where the base network is trained using statistical channel state information (CSI), and then fine-tuned based on instantaneous CSI. Finally, we propose a training and deployment framework for EMNN.
    \item %仿真结果验证了迁移学习方法加速EMNN训练的有效性。此外，我们对SIM (DPSIM)辅助E2E OFDM系统的性能进行了全面的评估，并将其与传统massive MIMO系统的性能进行了对比。
    The simulation results validate the effectiveness of the transfer learning approach in accelerating EMNN training. Furthermore, we provide a comprehensive performance evaluation of SIM (DPSIM)- assisted E2E OFDM systems and compare their performance with that of conventional massive MIMO systems.
\end{enumerate}

\subsection{Notations.}
We adopt bold lowercase and uppercase letters to denote vectors and matrices, respectively; $(\cdot)^*$, $(\cdot)^{\mathrm{T}}$, and $(\cdot)^{\mathrm{H}}$ represent the conjugate, transpose, and hermitian transpose, respectively; $|c|$, $\Re(c)$, and $\Im(c)$ refer to the magnitude, real part, and imaginary part, respectively, of a complex number $c$; $\|\cdot\|_{\text{F}}$ is the frobenius norm; $\|\cdot\|_{1}$ is the 1-norm; $\mathbb{E}(\cdot)$ stands for the expectation operation; $\odot$ denotes the hadamard product of matrices; $\operatorname{diag}(\mathbf{v})$ produces a diagonal matrix with the elements of $\mathbf{v}$ on the main diagonal; $\mathbb{C}^{x\times y}$ represents the space of $x\times y$ complex-valued matrices; $\mathbf{0}$ and $\mathbf{1}$ denote all-zero and all-one vectors, respectively, with appropriate dimensions, while $\mathbf{I}_{N}\in\mathbb{C}^{N\times N}$ denotes the identity matrix; the distribution of a circularly symmetric complex gaussian (CSCG) random vector with mean vector $\boldsymbol{\mu}$ and covariance matrix $\boldsymbol{\Sigma}\succeq\mathbf{0}$ is denoted by $\sim\mathcal{CN}(\boldsymbol{\mu},\boldsymbol{\Sigma})$.

\begin{figure*}[!t]
\centering
\includegraphics[width=7in]{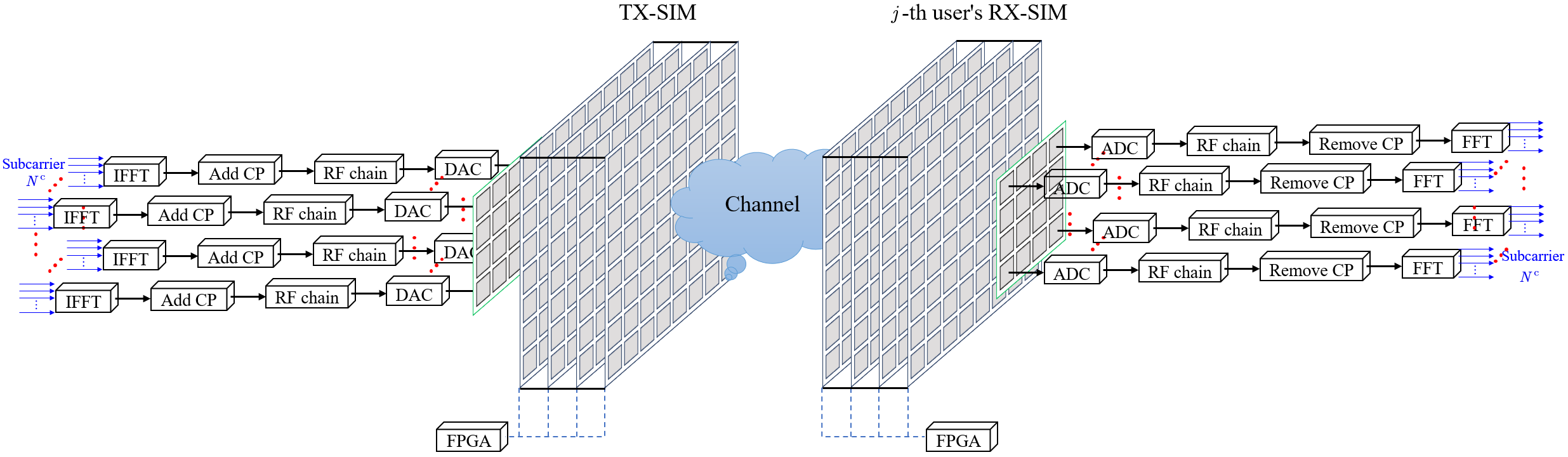}
\caption{Schematic diagram of the SIM-assisted E2E OFDM system.}
\label{Fig1}
\vspace{-0.2cm}
\end{figure*}

\subsection{Organization.}
%The rest of this paper is organized as follows. In Section~\ref{II},我们介绍了本文研究的系统模型，其中在~\ref{II.~A}和~\ref{II.~B}中依次介绍了SIM辅助E2E OFDM系统和单极化信道模型，在~\ref{II.~C}和~\ref{II.~D}中依次介绍了DPSIM辅助E2E OFDM系统和双极化信道模型。In Section~\ref{Ⅲ.~A}中，我们提出了SIM（DPSIM）辅助E2E OFDM系统的比特传输问题。在Section~\ref{Ⅲ.~B}到~\ref{Ⅲ.~E}中，我们依次介绍了EMNN的结构、模型训练和部署流程、复杂度分析、以及模型拓展性分析。在Section~\ref{Ⅴ}中给出本文结论之前，我们在Section~\ref{Ⅳ}中仿真验证了EMNN方法的有效性。
The remainder of this paper is organized as follows. Section~\ref{II} describes the system model considered in this work. Specifically, Sections~\ref{II.~A} and~\ref{II.~B} present the SIM-assisted E2E OFDM system and the single-polarization channel model, respectively, while Sections~\ref{II.~C} and~\ref{II.~D} introduce the DPSIM-assisted E2E OFDM system and the dual-polarization channel model, respectively. In Section~\ref{Ⅲ.~A}, the bit transmission problem for SIM (DPSIM)- assisted E2E OFDM systems is formulated. Sections~\ref{Ⅲ.~B} -- Sections~\ref{Ⅲ.~E} detail the EMNN architecture, model training and deployment procedures, complexity analysis, and scalability analysis, respectively. The effectiveness of the proposed EMNN approach is validated via simulations in Section~\ref{Ⅳ}, followed by concluding remarks in Section~\ref{Ⅴ}.

\section{System Model.} \label{II}

\begin{figure}[!t]
\centering
\includegraphics[width=3.5in]{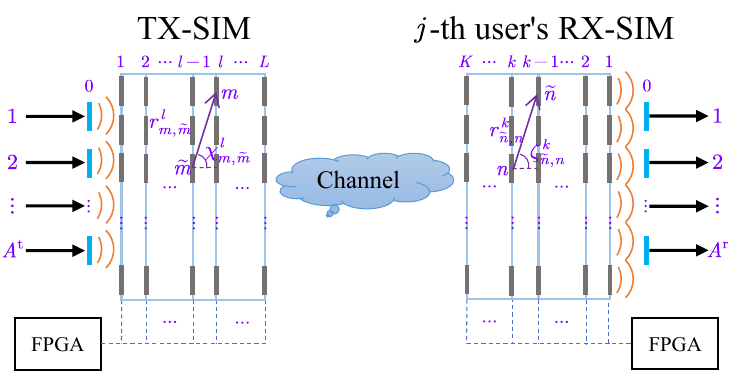}
\caption{TX-SIM and RX-SIM device parameter diagram.}
\label{Fig2}
\end{figure}

%在本节中，我们依次介绍了SIM辅助E2E OFDM系统，单极化信道模型，DPSIM辅助E2E OFDM系统和双极化信道模型。
This section sequentially presents the SIM-assisted E2E OFDM system, the single-polarization channel model, the DPSIM-assisted E2E OFDM system, and the dual-polarization channel model.

\subsection{SIM-assisted E2E OFDM system.} \label{II.~A}

% As shown in Fig. \ref{Fig1}, 我们考虑一个拥有$J$个用户的E2E OFDM系统，其中TX-SIM被集成在基站（BS）发射机中，RX-SIM被集成在每个用户设备（UE）的接收机中。$B$表示OFDM系统的带宽, $N_c$是OFDM系统中子载波的数量，$S$是独立数据流的数量。$L$ and $K$分别表示TX-SIM和RX-SIM的超表面层数。为便于表示，我们定义TX超表面的集合为$\mathcal{L} =\left\{0，1,2,\cdots,L\right\}$,第$j$个用户RX超表面的集合为\mathcal{K} _j=\left\{ 0,1,2,\cdots ,K \right\}，其中$0\in \mathcal{L}$ 和$0\in \mathcal{K} _j$分别代表BS的TX天线超表面和第$j$个用户的RX天线超表面。 为简化模型，我们假设UE中集成的RX-SIM器件和接收天线数量是一样的，构成SIM器件的所有超表面都是完全相同的。$A^{\mathrm{t}}=A^{\mathrm{tx},\mathrm{r}}\times A^{\mathrm{tx},\mathrm{c}}$ 和 $A^{\mathrm{r}}=A^{\mathrm{rx},\mathrm{r}}\times A^{\mathrm{rx},\mathrm{c}}$ 分别表示TX激活天线的数量和单个用户RX激活天线的数量.

As shown in Fig. \ref{Fig1}, we consider a $J$ user E2E OFDM system, where transmit SIM (TX-SIM) is integrated in the base station (BS) transmitter, and receive SIM (RX-SIM) is integrated in each user equipment (UE) receiver. To simplify the model, we assume that all UEs have the same number of receiving antennas and each integrates an identical RX-SIM device. Furthermore, all metasurface layers constituting a RX-SIM device are identical. Let $B$ and $N^{\mathrm{c}}$ denote the bandwidth and the number of subcarriers of the OFDM system, respectively. Moreover, as shown in Fig. \ref{Fig2}, $A^{\mathrm{t}}=A^{\mathrm{t}^{\mathrm{x}}}\times A^{\mathrm{t}^{\mathrm{y}}}$ and $A^{\mathrm{r}}=A^{\mathrm{r}^{\mathrm{x}}}\times A^{\mathrm{r}^{\mathrm{y}}}$ respectively denote the number of TX antennas and the number of RX antennas per UE. We further assume that TX-SIM comprises $L$ layers of metasurfaces, with $M=M^{\mathrm{x}}\times M^{\mathrm{y}}$ EM units on each layer, and one RX-SIM is equipped with $K$ layers of metasurfaces, with $N=N^{\mathrm{x}}\times N^{\mathrm{y}}$ EM units on each layer. For convenience of notation, we define the set of TX metasurfaces as $\mathcal{L} =\left\{0, 1, 2, \cdots, L\right\}$ and the set of RX metasurfaces for the $j$-th user as $\mathcal{K}_j = \left\{0, 1, 2, \cdots, K\right\}$, where $0 \in \mathcal{L}$ and $0 \in \mathcal{K}_j$ respectively represent the BS's TX antenna metasurface and the $j$-th user's RX antenna metasurface. The spacing between adjacent EM units in the TX-SIM and RX-SIM is denoted as $d^{\mathrm{t}}$ and $d^{\mathrm{r}}$, respectively.

\begin{figure}[!t]
\centering
\includegraphics[width=3.5in]{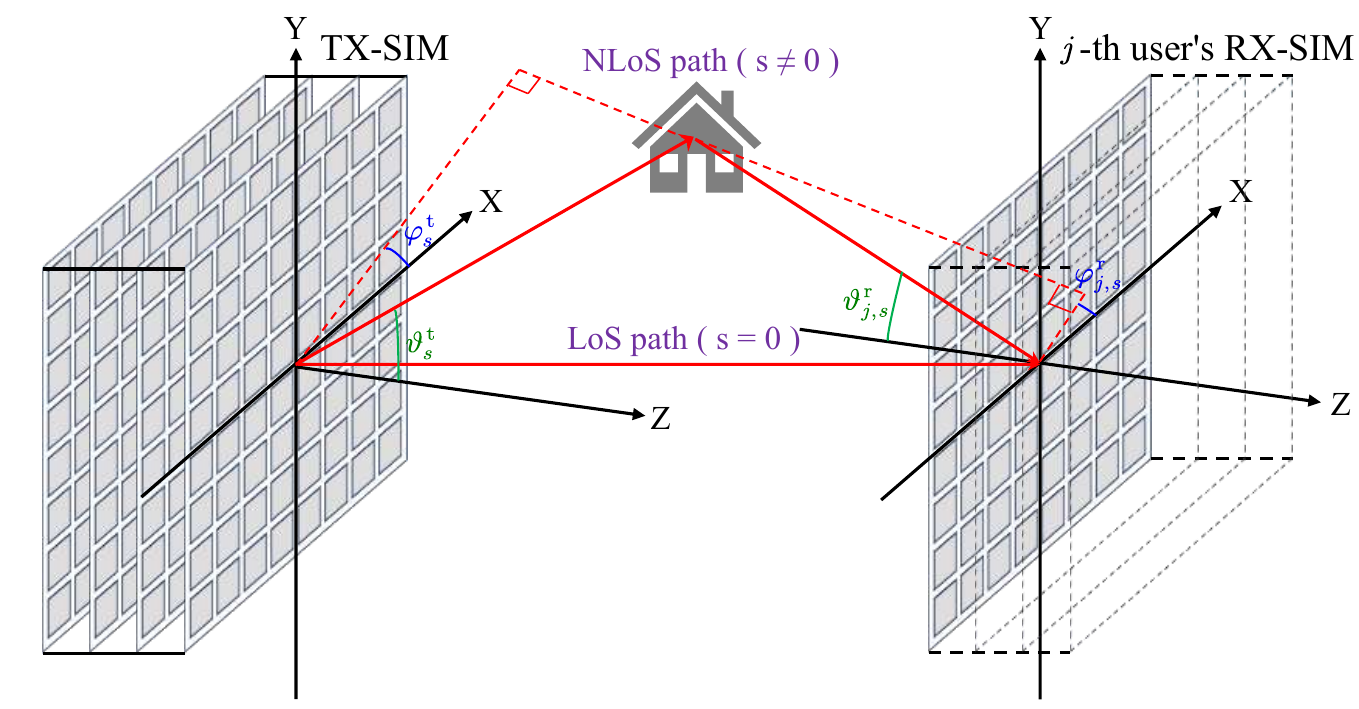}
\caption{Schematic diagram of multipath channel model.}
\label{Fig3}
\end{figure}

\begin{figure*}[!b]
\vspace{-0.7cm}
\begin{align*}
&\overline{\ \ \ \ \ \ \ \ \ \ \ \ \ \ \ \ \ \ \ \ \ \ \ \ \ \ \ \ \ \ \ \ \ \ \ \ \ \ \ \ \ \ \ \ \ \ \ \ \ \ \ \ \ \ \ \ \ \ \ \ \ \ \ \ \ \ \ \ \ \ \ \ \ \ \ \ \ \ \ \ \ \ \ \ \ \ \ \ \ \ \ \ \ \ \ \ \ \ \ \ \ \ \ \ \ \ \ \ \ \ \ \ \ \ \ \ \ \ \ \ \ \ \ \ \ \ \ \ \ \ \ \ \ \ \ \ \ \ \ \ \ \ \ \ \ \ \ }\\
\label{eq1}
&\ \ \ \ \ \ \ \ \ \ \ \ \ \ \ \ \mathbf{\Phi }^l=\mathrm{diag}\left( e^{\mathrm{j}\theta _{1}^{l}},e^{\mathrm{j}\theta _{2}^{l}},\cdots ,e^{\mathrm{j}\theta _{M}^{l}} \right) \in \mathbb{C} ^{M\times M},\theta _{m}^{l}\in \left[ 0,2\pi \right),m\in \mathcal{M},l\in \mathcal{L},\tag{1} \\
\label{eq2}
&\ \ \ \ \ \ \ \ \ \ \ \ \ \ \ \ \mathbf{\Psi }_{j}^{k}=\mathrm{diag}\left( e^{\mathrm{j}\xi _{1}^{j,k}},e^{\mathrm{j}\xi _{2}^{j,k}},\cdots ,e^{\mathrm{j}\xi _{N}^{j,k}} \right) \in \mathbb{C} ^{N\times N},\xi _{n}^{j,k}\in \left[ 0,2\pi \right) ,n\in \mathcal{N} ,k\in \mathcal{K} ,1\leqslant j\leqslant J.\tag{2}\\
&\overline{\ \ \ \ \ \ \ \ \ \ \ \ \ \ \ \ \ \ \ \ \ \ \ \ \ \ \ \ \ \ \ \ \ \ \ \ \ \ \ \ \ \ \ \ \ \ \ \ \ \ \ \ \ \ \ \ \ \ \ \ \ \ \ \ \ \ \ \ \ \ \ \ \ \ \ \ \ \ \ \ \ \ \ \ \ \ \ \ \ \ \ \ \ \ \ \ \ \ \ \ \ \ \ \ \ \ \ \ \ \ \ \ \ \ \ \ \ \ \ \ \ \ \ \ \ \ \ \ \ \ \ \ \ \ \ \ \ \ \ \ \ \ \ \ \ \ \ }\\
\label{eq3}
&\ \ \ \ \ \ \ \ \ \ \ \ \ \ \ \ \left[ \mathbf{V}_{i}^{l} \right] _{m,\tilde{m}}=\frac{S^{\mathrm{t}}\cos \chi _{m,\tilde{m}}^{l}}{r_{m,\tilde{m}}^{l}}\left( \frac{1}{2\pi r_{m,\tilde{m}}^{l}}-\mathrm{j}\frac{f_i}{c} \right) e^{\mathrm{j}2\pi r_{m,\tilde{m}}^{l}f_i/c},m\in \mathcal{M} ,\tilde{m}\in \mathcal{M} ,l\in \mathcal{L} ,1\leqslant i\leqslant N^{\mathrm{c}},\tag{3}\\
\label{eq4}
&\ \ \ \ \ \ \ \ \ \ \ \ \ \ \ \ \left[ \mathbf{U}_{i}^{k} \right] _{\tilde{n},n}=\frac{S^{\mathrm{r}}\cos \zeta _{\tilde{n},n}^{k}}{r_{\tilde{n},n}^{k}}\left( \frac{1}{2\pi r_{\tilde{n},n}^{k}}-\mathrm{j}\frac{f_i}{c} \right) e^{\mathrm{j}2\pi r_{\tilde{n},n}^{k}f_i/c},\tilde{n}\in \mathcal{N} ,n\in \mathcal{N} ,k\in \mathcal{K} ,1\leqslant i\leqslant N^{\mathrm{c}}.\tag{4}
\end{align*}
\end{figure*}

The transmission coefficient of the $l$-th layer for TX-SIM and the $k$-th layer for the $j$-th user's RX-SIM are defined as~\eqref{eq1} and~\eqref{eq2}, respectively. According to the Rayleigh-Sommerfeld diffraction theory~\cite{Lin2018Alloptical}, the transmission coefficient from the $\tilde{m}$-th EM unit on the $(l-1)$-th transmit metasurface layer to the $m$-th EM unit on the $l$-th transmit metasurface layer is expressed by~\eqref{eq3}, where $r_{m,\tilde{m}}^l$ denotes the corresponding transmission distance, $S^{\mathrm{t}}=d^{\mathrm{t}}\times d^{\mathrm{t}}$ is the area of each EM unit in the TX-SIM, while $\chi_{m,\tilde{m}}^l$ represents the angle between the propagation direction and the normal direction of the $(l-1)$-th transmit metasurface layer. $f_i$ denotes the frequency of the $i$-th subcarrier, and $c$ is the speed of light. Similarly, the transmission coefficient from the $n$-th EM unit on the $k$-th receive metasurface layer to the $\tilde{n}$-th EM unit on the $(k-1)$-th receive metasurface layer is expressed by~\eqref{eq4}, where $r_{\tilde{n},n}^k$ denotes the corresponding transmission distance, $S^{\mathrm{r}}=d^{\mathrm{r}}\times d^{\mathrm{r}}$ is the area of each EM unit in the RX-SIM, while $\zeta_{\tilde{n},n}^k$ represents the angle between the propagation direction and the normal direction of the $k$-th receive metasurface layer.

%信号在SIM中逐层传播所产生的累计效应可以通过矩阵连乘刻画。 对于TX-SIM，第$i$个子载波所对应的传播系数矩阵为
The cumulative effect generated by the layer-by-layer propagation of signals within SIM can be characterized by a chain of matrix products. For TX-SIM, the propagation coefficient matrix corresponding to the $i$-th subcarrier is
\begin{align*}
    \mathbf{T}_i=\mathbf{\Phi }^L\mathbf{V}_{i}^{L}\cdots \mathbf{\Phi }^2\mathbf{V}_{i}^{2}\mathbf{\Phi }^1\mathbf{V}_{i}^{1}\in \mathbb{C} ^{M\times A^{\mathrm{t}}}.\tag{5}
\end{align*}
%类似的，对于第j个UE的RX-SIM，其第$i$个子载波所对应的传播系数矩阵为
Similarly, for the RX-SIM of the $j$-th user, the propagation coefficient matrix corresponding to the $i$-th subcarrier is
\begin{align*}
    \mathbf{R}_{i,j}=\mathbf{U}_{i}^{1}\mathbf{\Psi }_{j}^{1}\mathbf{U}_{i}^{2}\mathbf{\Psi }_{j}^{2}\cdots \mathbf{U}_{i}^{K}\mathbf{\Psi }_{j}^{K}\in \mathbb{C} ^{A^{\mathrm{r}}\times N}.\tag{6}
\end{align*}

\begin{figure*}[t]
\centering
\includegraphics[width=7in]{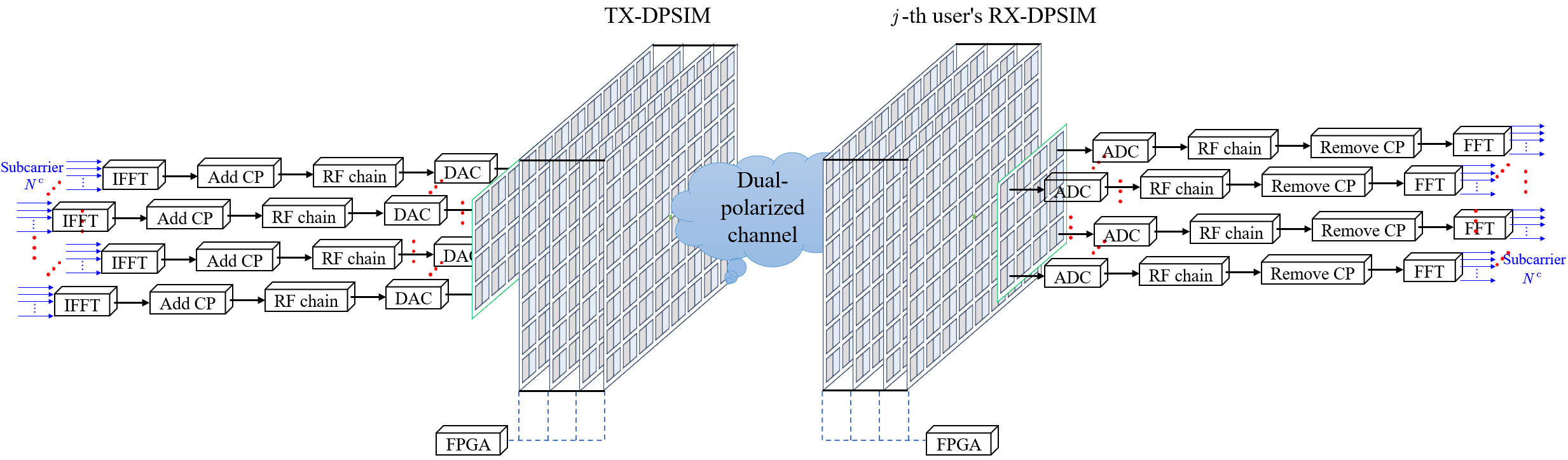}
\caption{Schematic diagram of the DPSIM-assisted E2E OFDM system.}
\label{Fig4}
\vspace{-0.2cm}
\end{figure*}

\begin{figure*}[!b]
\vspace{-0.7cm}
\begin{align*}
&\overline{\ \ \ \ \ \ \ \ \ \ \ \ \ \ \ \ \ \ \ \ \ \ \ \ \ \ \ \ \ \ \ \ \ \ \ \ \ \ \ \ \ \ \ \ \ \ \ \ \ \ \ \ \ \ \ \ \ \ \ \ \ \ \ \ \ \ \ \ \ \ \ \ \ \ \ \ \ \ \ \ \ \ \ \ \ \ \ \ \ \ \ \ \ \ \ \ \ \ \ \ \ \ \ \ \ \ \ \ \ \ \ \ \ \ \ \ \ \ \ \ \ \ \ \ \ \ \ \ \ \ \ \ \ \ \ \ \ \ \ \ \ \ \ \ \ \ \ }\\
\label{eq10}
&\ \ \ \ \ \ \ \ \ \ \ \ \ \ \ \ \ \ \ \ \ \ \ \boldsymbol{\alpha }_{i,s}^{\mathrm{t}}=\boldsymbol{\alpha }_{s}^{\mathrm{t}^{\mathrm{x}}}\left( \frac{2\mathrm{\pi c}}{f_i}d^{\mathrm{t}}\sin\mathrm{(}\vartheta _{s}^{\mathrm{t}})\cos\mathrm{(}\varphi _{s}^{\mathrm{t}}) \right) \otimes \boldsymbol{\alpha }_{s}^{\mathrm{t}^{\mathrm{y}}}\left( \frac{2\mathrm{\pi c}}{f_i}d^{\mathrm{t}}\sin\mathrm{(}\vartheta _{s}^{\mathrm{t}})\sin\mathrm{(}\varphi _{s}^{\mathrm{t}}) \right) 
,\tag{10} \\
\label{eq11}
&\ \ \ \ \ \ \ \ \ \ \ \ \ \ \ \ \ \ \ \ \ \ \ \boldsymbol{\alpha }_{i,j,s}^{\mathrm{r}}=\boldsymbol{\alpha }_{j,s}^{\mathrm{r}^{\mathrm{x}}}\left( \frac{2\mathrm{\pi c}}{f_i}d^{\mathrm{r}}\sin\mathrm{(}\vartheta _{j,s}^{\mathrm{r}})\cos\mathrm{(}\varphi _{j,s}^{\mathrm{r}}) \right) \otimes \boldsymbol{\alpha }_{j,s}^{\mathrm{r}^{\mathrm{y}}}\left( \frac{2\mathrm{\pi c}}{f_i}d^{\mathrm{r}}\sin\mathrm{(}\vartheta _{j,s}^{\mathrm{r}})\sin\mathrm{(}\varphi _{j,s}^{\mathrm{r}}) \right) .\tag{11}\\
&\overline{\ \ \ \ \ \ \ \ \ \ \ \ \ \ \ \ \ \ \ \ \ \ \ \ \ \ \ \ \ \ \ \ \ \ \ \ \ \ \ \ \ \ \ \ \ \ \ \ \ \ \ \ \ \ \ \ \ \ \ \ \ \ \ \ \ \ \ \ \ \ \ \ \ \ \ \ \ \ \ \ \ \ \ \ \ \ \ \ \ \ \ \ \ \ \ \ \ \ \ \ \ \ \ \ \ \ \ \ \ \ \ \ \ \ \ \ \ \ \ \ \ \ \ \ \ \ \ \ \ \ \ \ \ \ \ \ \ \ \ \ \ \ \ \ \ \ \ }\\
\label{eq16}
&\ \ \ \ \mathbf{\Phi }_{p}^{l}=\mathrm{diag}\left( e^{\mathrm{j}\theta _{p,1}^{l}},e^{\mathrm{j}\theta _{p,2}^{l}},\cdots ,e^{\mathrm{j}\theta _{p,M^{\mathrm{dp}}}^{l}} \right) \in \mathbb{C} ^{M^{\mathrm{dp}}\times M^{\mathrm{dp}}},\theta _{p,m}^{l}\in \left[ 0,2\pi \right) ,m\in \mathcal{M} ^{\mathrm{dp}},l\in \mathcal{L} ^{\mathrm{dp}},p\in \left\{ 0,1 \right\}, \tag{16}\\
\label{eq17}
&\ \ \ \ \mathbf{\Psi }_{j,p}^{k}=\mathrm{diag}\left( e^{\mathrm{j}\xi _{p,1}^{j,k}},e^{\mathrm{j}\xi _{p,2}^{j,k}},\cdots ,e^{\mathrm{j}\xi _{p,N^{\mathrm{dp}}}^{j,k}} \right) \in \mathbb{C} ^{N^{\mathrm{dp}}\times N^{\mathrm{dp}}},\xi _{p,n}^{j,k}\in \left[ 0,2\pi \right) ,n\in \mathcal{N} ^{\mathrm{dp}},k\in \mathcal{K} ^{\mathrm{dp}},p\in \left\{ 0,1 \right\} ,1\leqslant j\leqslant J.\tag{17}\\
\end{align*}
\end{figure*}

%综上，第$j$个用户在第$i$个子载波频段接受到的信号为
In summary, the signal received by the $j$-th user on the $i$-th subcarrier is given by
\begin{align*}
\mathbf{y}_{i,j}=\mathbf{R}_{i,j}\mathbf{G}_{i,j}\mathbf{T}_{i}\mathbf{p}_{i}\odot\mathbf{x}_{i}+\mathbf{n}_{i,j},\tag{7}
\end{align*}
where $\mathbf{n}_{i,j}\in \mathbb{C} ^{A^{\mathrm{r}}\times 1}$ is the receiver noise vector with distribution $ \mathcal{C} \mathcal{N} \left( 0,\sigma ^2\mathbf{I}_{A^{\mathrm{r}}} \right) $, $\mathbf{p}_i\in \mathbb{C} ^{A^{\mathrm{t}}\times 1}$ denotes the vector of transmit powers, $\mathbf{G}_{i,j}$ is the channel coefficient matrix between the $L$-th layer metasurface of the TX-SIM and the $K$-th layer metasurface of the $j$-th user's RX-SIM, $\mathbf{x}_i\in \mathbb{C} ^{A^{\mathrm{t}}\times 1}$ is the signal vector satisfying $\mathbb{E} \left\{ \mathbf{xx}^{\mathrm{H}} \right\} =\mathbf{I}_{A^{\mathrm{t}}N^{\mathrm{c}}}$, with $\mathbf{x}=\left[ \mathbf{x}_{1}^{\mathrm{T}},\mathbf{x}_{2}^{\mathrm{T}},\cdots ,\mathbf{x}_{N^{\mathrm{c}}}^{\mathrm{T}} \right] ^{\mathrm{T}}
$.

%Let $N^{\mathrm{bit}}$ is the number of bits corresponding to an OFDM symbol transmitted by the BS, $N_{j}^{\mathrm{bit}}$表示第$j$个用户在一个OFDM符号中所对应的接收比特数量。$\mathbf{b}\in \left\{ 0,1 \right\} ^{N^{\mathrm{bit}}\times 1}$和$\hat{\mathbf{b}}_j\in \left\{ 0,1 \right\} ^{N_{j}^{\mathrm{bit}}\times 1}$分别为发送比特向量和接受比特向量。 则SIM辅助E2E OFDM系统的比特传输模型为
Let $N^{\mathrm{bit}}$ denote the number of bits corresponding to an OFDM symbol transmitted by the BS, and $N_{j}^{\mathrm{bit}}$ represent the number of bits received by the $j$-th user within one OFDM symbol. $\mathbf{b} \in \{0,1\}^{N^{\mathrm{bit}}\times 1}$ and $\hat{\mathbf{b}}_j \in \{0,1\}^{N_{j}^{\mathrm{bit}}\times 1}$ denote the transmitted bit vector and the $j$-th user's received bit vector, respectively. Then, the bit transmission model of SIM-assisted E2E OFDM system model can be expressed as
\begin{align*}
\label{eq8}
\hat{\mathbf{b}}_j=\mathbf{f}_{j}^{\mathrm{D}}\left( \left[ \begin{array}{c}
	\mathbf{R}_{1,j}\mathbf{G}_{1,j}\mathbf{T}_1\mathbf{p}_1\odot \mathbf{f}_{1}^{\mathrm{M}}(\mathbf{b})+\mathbf{n}_{1,j}\\
	\mathbf{R}_{2,j}\mathbf{G}_{2,j}\mathbf{T}_2\mathbf{p}_2\odot \mathbf{f}_{2}^{\mathrm{M}}(\mathbf{b})+\mathbf{n}_{2,j}\\
	\vdots\\
	\mathbf{R}_{N^{\mathrm{c}},j}\mathbf{G}_{N^{\mathrm{c}},j}\mathbf{T}_{N^{\mathrm{c}}}\mathbf{p}_{N^{\mathrm{c}}}\odot \mathbf{f}_{N^{\mathrm{c}}}^{\mathrm{M}}(\mathbf{b})+\mathbf{n}_{N^{\mathrm{c}},j}\\
\end{array} \right] \right) ,\tag{8}
\end{align*}
%其中$\mathbf{x}_i=\mathbf{f}_{i}^{\mathrm{M}}(\mathbf{b})$, $i=1,2,\cdots,N^{\mathrm{c}}$表示从$\mathbf{b}$到第$i$个子载波信号的调制映射，$\hat{\mathbf{b}}_j=\mathbf{f}_{j}^{\mathrm{dem}}\left( \left[ \mathbf{y}_{1,j},\mathbf{y}_{2,j},\cdots ,\mathbf{y}_{N^{\mathrm{c}},j} \right] ^{\mathrm{T}} \right) $表示第$j$个用户接收的多载波信号到接收比特向量$\hat{\mathbf{b}}_j\in \left\{ 0,1 \right\} ^{N^{\mathrm{bit}}\times 1}$的解调映射。
where $\mathbf{x}_i=\mathbf{f}_{i}^{\mathrm{M}}(\mathbf{b})$, $i=1,2,\cdots,N^{\mathrm{c}}$ denotes the modulation mapping from $\mathbf{b}$ to the signal on the $i$-th subcarrier. Furthermore, $\hat{\mathbf{b}}_j=\mathbf{f}_{j}^{\mathrm{D}}\left( \left[ \mathbf{y}_{1,j},\mathbf{y}_{2,j},\cdots,\mathbf{y}_{N^{\mathrm{c}},j} \right] ^{\mathrm{T}} \right) $, $j=1,2,\cdots,J$ is the demodulation mapping of the received wideband signal at the $j$-th user to $\hat{\mathbf{b}}_j$.

\subsection{Single-polarization channel model (SIM-assisted E2E OFDM system channel model).} \label{II.~B}

As shown in Fig. \ref{Fig3}, to precisely model the spatial and frequency-selective properties inherent in wideband propagation, the proposed SIM-assisted E2E OFDM system utilizes a geometric multi-path channel model. There are $S$ scatterers between the BS and UEs, each of which affects the signal propagation path. Let $s=0$ denote the line-of-sight (LoS) path and $s=1,2,\cdots,S$ denote non-LoS (NLoS) paths. We assume that the gain and delay of the $i$-th subcarrier from the BS to the $j$-th user through the $s$-th path are $g_{i,j,s}$ and $\tau_{i,j,s}$, respectively. Although the transmission signals between TX and RX are time-domain signals, the wideband channel model is often represented in the frequency-domain form to simplify the IFFT-FFT progress in the beamspace~\cite{7400949}. For the $i$-th subcarrier with frequency $f_i$, the channel coefficient matrix $\mathbf{G}_{i,j} \in \mathbb{C}^{N \times M}$ is modeled as
\begin{align*}
    \label{eq9}
    \mathbf{G}_{i,j}=\sum_{s=0}^S{g_{i,j,s}}e^{-\mathrm{j}2\pi f_i\tau _{i,j,s}}\boldsymbol{\alpha }_{i,j,s}^{\mathrm{r}}\left( \boldsymbol{\alpha }_{i,s}^{\mathrm{t}} \right) ^{\mathrm{H}},\tag{9}
\end{align*}
where $\boldsymbol{\alpha }_{i,s}^{\mathrm{t}} \in \mathbb{C}^{M \times 1}$ and $\boldsymbol{\alpha }_{i,j,s}^{\mathrm{r}} \in \mathbb{C}^{N \times 1}$ are the corresponding transmit and receive steering vectors, respectively.
%其中$\mathbf{\alpha }_{s,i}^{\mathrm{t}} \in \mathbb{C}^{M \times 1}$ and $\mathbf{\alpha }_{j,s,i}^{\mathrm{r}} \in \mathbb{C}^{N \times 1}$分别为对应的发射和接收导向矢量。%第$s$条传播路径相对于TX-SIM的俯仰角和方位角分别为$\vartheta_s^{\mathrm{t}} \in [0, \pi/2)$和$\varphi _{s}^{\mathrm{t}}\in [0,2\pi )$，相对于UE$j$的俯仰角和方位角分别为$\vartheta_{j,s}^{\mathrm{r}} \in [0, \pi/2)$ and $\varphi_{j,s}^{\mathrm{r}} \in [0, 2\pi)$。The center of the TX-SIM is located at the origin $(0, 0, 0)$.

Without loss of generality, we assume that the TX-SIM and RX-SIM are vertically aligned in a 3D cartesian coordinate system. For the $s$-th propagation path, its elevation and azimuth angles relative to the TX-SIM are $\vartheta_s^{\mathrm{t}} \in [0, \pi/2)$ and $\varphi _{s}^{\mathrm{t}}\in [0,2\pi )$, respectively, while those relative to the $j$-th user's RX-SIM are $\vartheta_{j,s}^{\mathrm{r}} \in [0, \pi/2)$ and $\varphi_{j,s}^{\mathrm{r}} \in [0, 2\pi)$. The steering vector $\boldsymbol{\alpha }_{i,s}^{\mathrm{t}}$ and $\boldsymbol{\alpha }_{i,j,s}^{\mathrm{r}}$ can be written as~\eqref{eq10} and~\eqref{eq11}, respectively, where $\boldsymbol{\alpha }_{s}^{\mathrm{t}^{\mathrm{x}}}(*)\in \mathbb{C} ^{M^{\mathrm{x}}\times 1}$ and $\boldsymbol{\alpha }_{s}^{\mathrm{t}^{\mathrm{y}}}(*)\in \mathbb{C} ^{M^{\mathrm{y}}\times 1}$ are defined as follows
\begin{align*}
\label{eq12}
&[\boldsymbol{\alpha }_{s}^{\mathrm{t}^{\mathrm{x}}}(*)]_{m^{\mathrm{x}}}\triangleq e^{\mathrm{j(}m^{\mathrm{x}}-1)*},m^{\mathrm{x}}=1,2,\cdots ,M^{\mathrm{x}},\tag{12} \\
\label{eq13}
&[\boldsymbol{\alpha }_{s}^{\mathrm{t}^{\mathrm{y}}}(*)]_{m^{\mathrm{y}}}\triangleq e^{\mathrm{j(}m^{\mathrm{y}}-1)*},m^{\mathrm{y}}=1,2,\cdots ,M^{\mathrm{y}},\tag{13}
\end{align*}
where $\mathbf{\alpha }_{s}^{\mathrm{r}^{\mathrm{x}}}(*)\in \mathbb{C} ^{N^{\mathrm{x}}\times 1}$ and $\mathbf{\alpha }_{s}^{\mathrm{r}^{\mathrm{y}}}(*)\in \mathbb{C} ^{N^{\mathrm{y}}\times 1}$ are defined as
\begin{align*}
\label{eq14}
&[\boldsymbol{\alpha }_{j,s}^{\mathrm{r}^{\mathrm{x}}}(*)]_{n^{\mathrm{x}}}\triangleq e^{\mathrm{j(}n^{\mathrm{x}}-1)*},n^{\mathrm{x}}=1,2,\cdots ,N^{\mathrm{x}},\tag{14} \\
\label{eq15}
&[\boldsymbol{\alpha }_{j,s}^{\mathrm{r}^{\mathrm{y}}}(*)]_{n^{\mathrm{y}}}\triangleq e^{\mathrm{j(}n^{\mathrm{y}}-1)*},n^{\mathrm{y}}=1,2,\cdots ,N^{\mathrm{y}}.\tag{15}
\end{align*}

\begin{figure}[!t]
\centering
\includegraphics[width=3.5in]{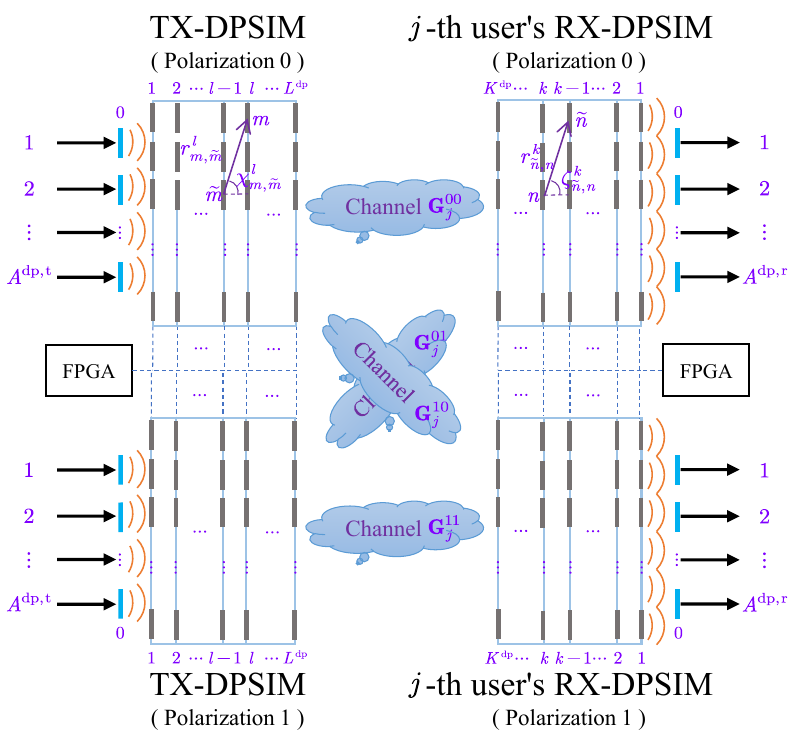}
\caption{Schematic diagram of the parameters of TX-DPSIM and RX-DPSIM in the two orthogonal polarization directions.}
\label{Fig5}
\end{figure}

\subsection{DPSIM-assisted E2E OFDM system.} \label{II.~C}

%与Fig. \ref{Fig1}类似，Fig. \ref{Fig4}展示了一个DPSIM辅助的E2E OFDM系统，其中TX-DPSIM和RX-DPSIM分别被集成在收发两侧来在波域对信号进行处理。Considering that each layer of the metasurface in DPSIM can independently control the phase for different polarization directions, therefore DPSIM can be equivalently treated as two identical and isolated single-polarization SIMs. 
Similar to Fig.~\ref{Fig1}, Fig.~\ref{Fig4} illustrates a DPSIM-assisted E2E OFDM system, where the TX-DPSIM and RX-DPSIM are integrated at the transmitter and receiver sides, respectively, to process the signal in the wave domain. Considering that each layer of the metasurface in DPSIM can independently control the phase for different polarization directions, the DPSIM can thus be equivalently regarded as two identical and isolated single-polarization SIMs. To facilitate understanding the model, Fig.~\ref{Fig5} presents a schematic of the HMIMO system with two orthogonal polarization directions, where we let $p \in \{0,1\}$ denote both polarizations. Consistent with the assumptions in Section~\ref{II.~A}, each UE integrates an identical RX-DPSIM device, and each metasurface layer constituting the RX-DPSIM is perfectly identical. Moreover, $A^{\mathrm{dp},\mathrm{t}}=A^{\mathrm{dp},\mathrm{t}^{\mathrm{x}}}\times A^{\mathrm{dp},\mathrm{t}^{\mathrm{y}}}$ and $A^{\mathrm{dp},\mathrm{r}}=A^{\mathrm{dp},\mathrm{r}^{\mathrm{x}}}\times A^{\mathrm{dp},\mathrm{r}^{\mathrm{y}}}$ respectively denote the number of TX dual-polarized antennas and the number of RX dual-polarized antennas per UE. We further assume that TX-DPSIM comprises $L^{\mathrm{dp}}$ layers of metasurfaces, with $M^{\mathrm{dp}}=M^{\mathrm{dp},\mathrm{x}}\times M^{\mathrm{dp},\mathrm{y}}$ DPEM units on each layer, and one RX-DPSIM is equipped with $K^{\mathrm{dp}}$ layers of metasurfaces, with $N^{\mathrm{dp}}=N^{\mathrm{dp},\mathrm{x}}\times N^{\mathrm{dp},\mathrm{y}}$ DPEM units on each layer. For convenience of notation, we define the set of TX metasurfaces as $\mathcal{L} ^{\mathrm{dp}}=\left\{ 0,1,2,\cdots ,L^{\mathrm{dp}} \right\} $ and the set of RX metasurfaces for the $j$-th user as $\mathcal{K} _{j}^{\mathrm{dp}}=\left\{ 0,1,2,\cdots ,K^{\mathrm{dp}} \right\}$, where $0 \in \mathcal{L}^{\mathrm{dp}}$ and $0 \in \mathcal{K} _{j}^{\mathrm{dp}}$ respectively represent the BS's TX antenna metasurface and the $j$-th user's RX antenna metasurface. The spacing between adjacent DPEM units in the TX-DPSIM is denoted as $d^{\mathrm{dp},\mathrm{t}}$, and the spacing in the RX-DPSIM is denoted as $d^{\mathrm{dp},\mathrm{r}}$.

The transmission coefficient of the $l$-th layer for TX-DPSIM and the $k$-th layer for the $j$-th user's RX-DPSIM are defined as~\eqref{eq16} and~\eqref{eq17}, respectively. Considering that the interaction between EM waves and environmental scatterers is the main mechanism for changing their polarization state~\cite{5979177}, we assume that the polarization state of the EM waves remains unchanged during their propagation between DPSIM layers. Therefore, the overall coefficient matrices for TX-DPSIM and the $j$-th user's RX-DPSIM are respectively given by~\eqref{eq18} and~\eqref{eq19}, where $\mathbf{V}_{i}^{l},\ l\in \mathcal{L} ^{\mathrm{dp}}$ and $\mathbf{U}_{i}^{k},\ k\in \mathcal{K} ^{\mathrm{dp}}$ are calculated in the similar way as in Section~\ref{II.~A} for \eqref{eq3} and \eqref{eq4}.

\begin{figure*}[b]
\vspace{-0.7cm}
\begin{align*}
\label{eq18}
&\overline{\ \ \ \ \ \ \ \ \ \ \ \ \ \ \ \ \ \ \ \ \ \ \ \ \ \ \ \ \ \ \ \ \ \ \ \ \ \ \ \ \ \ \ \ \ \ \ \ \ \ \ \ \ \ \ \ \ \ \ \ \ \ \ \ \ \ \ \ \ \ \ \ \ \ \ \ \ \ \ \ \ \ \ \ \ \ \ \ \ \ \ \ \ \ \ \ \ \ \ \ \ \ \ \ \ \ \ \ \ \ \ \ \ \ \ \ \ \ \ \ \ \ \ \ \ \ \ \ \ \ \ \ \ \ \ \ \ \ \ \ \ \ \ \ \ \ \ }\\
&\ \ \ \ \ \ \ \ \ \mathbf{T}_{i}^{\mathrm{dp}}=\left[ \begin{matrix}
	\mathbf{\Phi }_{0}^{L}&		\mathbf{0}\\
	\mathbf{0}&		\mathbf{\Phi }_{1}^{L}\\
\end{matrix} \right] \left[ \begin{matrix}
	\mathbf{V}_{i}^{L}&		\mathbf{0}\\
	\mathbf{0}&		\mathbf{V}_{i}^{L}\\
\end{matrix} \right] \cdots \left[ \begin{matrix}
	\mathbf{\Phi }_{0}^{2}&		\mathbf{0}\\
	\mathbf{0}&		\mathbf{\Phi }_{1}^{2}\\
\end{matrix} \right] \left[ \begin{matrix}
	\mathbf{V}_{i}^{2}&		\mathbf{0}\\
	\mathbf{0}&		\mathbf{V}_{i}^{2}\\
\end{matrix} \right] \left[ \begin{matrix}
	\mathbf{\Phi }_{0}^{1}&		\mathbf{0}\\
	\mathbf{0}&		\mathbf{\Phi }_{1}^{1}\\
\end{matrix} \right] \left[ \begin{matrix}
	\mathbf{V}_{i}^{1}&		\mathbf{0}\\
	\mathbf{0}&		\mathbf{V}_{i}^{1}\\
\end{matrix} \right] \in \mathbb{C} ^{2M^{\mathrm{dp}}\times 2A^{\mathrm{dp},\mathrm{t}}},\tag{18}\\
\label{eq19}
&\ \ \ \ \ \ \ \ \ \mathbf{R}_{i,j}^{\mathrm{dp}}=\left[ \begin{matrix}
	\mathbf{U}_{i}^{1}&		\mathbf{0}\\
	\mathbf{0}&		\mathbf{U}_{i}^{1}\\
\end{matrix} \right] \left[ \begin{matrix}
	\mathbf{\Psi }_{j,0}^{1}&		\mathbf{0}\\
	\mathbf{0}&		\mathbf{\Psi }_{j,1}^{1}\\
\end{matrix} \right] \left[ \begin{matrix}
	\mathbf{U}_{i}^{2}&		\mathbf{0}\\
	\mathbf{0}&		\mathbf{U}_{i}^{2}\\
\end{matrix} \right] \left[ \begin{matrix}
	\mathbf{\Psi }_{j,0}^{2}&		\mathbf{0}\\
	\mathbf{0}&		\mathbf{\Psi }_{j,1}^{2}\\
\end{matrix} \right] \cdots \left[ \begin{matrix}
	\mathbf{U}_{i}^{K}&		\mathbf{0}\\
	\mathbf{0}&		\mathbf{U}_{i}^{K}\\
\end{matrix} \right] \left[ \begin{matrix}
	\mathbf{\Psi }_{j,0}^{K}&		\mathbf{0}\\
	\mathbf{0}&		\mathbf{\Psi }_{j,1}^{K}\\
\end{matrix} \right] \in \mathbb{C} ^{2A^{\mathrm{dp},\mathrm{r}}\times 2N^{\mathrm{dp}}}.\tag{19}
\end{align*}
\end{figure*}

In summary, the signal received by the $j$-th user on the $i$-th subcarrier is given by
\begin{align*}
\label{eq20}
\mathbf{y}_{i,j}^{\mathrm{dp}}=\mathbf{R}_{i,j}^{\mathrm{dp}}\mathbf{G}_{i,j}^{\mathrm{dp}}\mathbf{T}_{i}^{\mathrm{dp}}\mathbf{p}_{i}^{\mathrm{dp}}\odot \mathbf{x}_{i}^{\mathrm{dp}}+\mathbf{n}_{i,j}^{\mathrm{dp}},\tag{20}
\end{align*}
where $\mathbf{n}_{i,j}^{\mathrm{dp}}\in \mathbb{C} ^{2A^{\mathrm{dp},\mathrm{r}}\times 1}$ is the receiver noise vector with distribution $\mathcal{C} \mathcal{N} (0,\sigma ^2\mathbf{I}_{2A^{\mathrm{dp},\mathrm{r}}})$, $\mathbf{p}_{i}^{\mathrm{dp}}\in \mathbb{C} ^{2A^{\mathrm{dp},\mathrm{t}}\times 1}$ denotes the vector of transmit powers, $\mathbf{G}_{i,j}^{\mathrm{dp}}$ is the channel coefficient matrix between the $L^{\mathrm{dp}}$-th layer metasurface of the TX-DPSIM and the $K^{\mathrm{dp}}$-th layer metasurface of the $j$-th user's RX-DPSIM, $\mathbf{x}_{i}^{\mathrm{dp}}\in \mathbb{C} ^{2A^{\mathrm{dp},\mathrm{t}}\times 1}$ is the signal vector satisfying $\mathbb{E} \left\{ \mathbf{x}^{\mathrm{dp}}\left( \mathbf{x}^{\mathrm{dp}} \right) ^{\mathrm{H}} \right\} =\mathbf{I}_{2A^{\mathrm{dp},\mathrm{t}}N^{\mathrm{c}}}$, with $\mathbf{x}^{\mathrm{dp}}=\left[ \left( \mathbf{x}_{1}^{\mathrm{dp}} \right) ^{\mathrm{T}},\left( \mathbf{x}_{2}^{\mathrm{dp}} \right) ^{\mathrm{T}},\cdots ,\left( \mathbf{x}_{N^{\mathrm{c}}}^{\mathrm{dp}} \right) ^{\mathrm{T}} \right] ^{\mathrm{T}}$.

%与Section~\ref{II.~A}中所介绍的类似，DPSIM辅助的端到端HMIMO OFDM系统模型为
Analogous to the system described in Section~\ref{II.~A}, the bit transmission model of the DPSIM-assisted E2E OFDM system is given by
\begin{align*}
\hat{\mathbf{b}}_j=\mathbf{f}_{j}^{\mathrm{dp},\mathrm{D}}\left( \left[ \begin{array}{c}
	\mathbf{R}_{1,j}^{\mathrm{dp}}\mathbf{G}_{1,j}^{\mathrm{dp}}\mathbf{T}_{1}^{\mathrm{dp}}\mathbf{p}_{1}^{\mathrm{dp}}\odot \mathbf{f}_{1}^{\mathrm{dp},\mathrm{M}}(\mathbf{b})+\mathbf{n}_{1,j}^{\mathrm{dp}}\\
	\mathbf{R}_{2,j}^{\mathrm{dp}}\mathbf{G}_{2,j}^{\mathrm{dp}}\mathbf{T}_{2}^{\mathrm{dp}}\mathbf{p}_{2}^{\mathrm{dp}}\odot \mathbf{f}_{2}^{\mathrm{dp},\mathrm{M}}(\mathbf{b})+\mathbf{n}_{2,j}^{\mathrm{dp}}\\
	\vdots\\
	\mathbf{R}_{N^{\mathrm{c}},j}^{\mathrm{dp}}\mathbf{G}_{N^{\mathrm{c}},j}^{\mathrm{dp}}\mathbf{T}_{N^{\mathrm{c}}}^{\mathrm{dp}}\mathbf{p}_{N^{\mathrm{c}}}^{\mathrm{dp}}\odot \mathbf{f}_{N^{\mathrm{c}}}^{\mathrm{dp},\mathrm{M}}(\mathbf{b})+\mathbf{n}_{N^{\mathrm{c}},j}^{\mathrm{dp}}\\
\end{array} \right] \right) ,\tag{21}
\end{align*}
%其中$\mathbf{x}_i=\mathbf{f}_{i}^{\mathrm{mod}}(\mathbf{b}),i=1,2,\cdots N^{\mathrm{c}}$表示发送比特向量到第$i$个子载波的调制映射，$\hat{\mathbf{b}}_j=\mathbf{f}_{j}^{\mathrm{dem}}\left( \left[ \mathbf{y}_{1,j},\mathbf{y}_{2,j},\cdots ,\mathbf{y}_{N^{\mathrm{c}},j} \right] ^{\mathrm{T}} \right) $表示第$j$个用户接收的宽带信号到接收比特向量$\hat{\mathbf{b}}_j\in \left\{ 0,1 \right\} ^{N^{\mathrm{bit}}\times 1}$的解调映射。
where $\mathbf{x}_{i}^{\mathrm{dp}}=\mathbf{f}_{i}^{\mathrm{dp},\mathrm{M}}(\mathbf{b})$, $i=1,2,\cdots,N^{\mathrm{c}}$ denotes the modulation mapping from $\mathbf{b}$ to the singnal on the $i$-th subcarrier. Furthermore, $\hat{\mathbf{b}}_j=\mathbf{f}_{j}^{\mathrm{dp},\mathrm{D}}\left( \left[ \mathbf{y}_{1,j}^{\mathrm{dp}},\mathbf{y}_{2,j}^{\mathrm{dp}},\cdots ,\mathbf{y}_{N^{\mathrm{c}},j}^{\mathrm{dp}} \right] ^{\mathrm{T}} \right) $, $j=1,2,\cdots,J$ is the demodulation mapping of the received wideband signal at the $j$-th user to $\hat{\mathbf{b}}_j$.

\subsection{Dual-polarization channel model (DPSIM-assisted E2E OFDM system channel model) .} \label{II.~D}

%Due to the presence of dual-polarization defects (polarization conversion), we use four matrices ( 00, 11, 10, and 01 ) to characterize the channel. These matrices respectively represent the channel from polarization 0 to 0, from polarization 1 to 1, from polarization 0 to 1, and from polarization 1 to 0~\cite{9497725}. 此外，我们引入cross-polarization discrimination (XPD)来刻画信道的交叉极化干扰强度，并假设基站TX-DPSIM到所有用户RX-DPSIM之间信道的XPD是相同的。Then
Due to the presence of dual-polarization defects (polarization conversion), we use four matrices ( 00, 11, 10, and 01 ) to characterize the channel. These matrices respectively represent the channel from polarization 0 to 0, from polarization 1 to 1, from polarization 0 to 1, and from polarization 1 to 0~\cite{9497725}. In addition, we introduce the cross-polarization discrimination (XPD) metric to quantify the strength of cross-polarization interference, and assume that the XPD value between the BS's TX-DPSIM and the RX-DPSIM of all users is identical. Then,
\begin{align*}
\label{eq22}
\mathrm{XPD}=\frac{\mathbb{E} \{\parallel \mathbf{G}_{i,j}^{00}\parallel ^2\}}{\mathbb{E} \{\parallel \mathbf{G}_{i,j}^{10}\parallel ^2\}}=\frac{\mathbb{E} \{\parallel \mathbf{G}_{i,j}^{11}\parallel ^2\}}{\mathbb{E} \{\parallel \mathbf{G}_{i,j}^{01}\parallel ^2\}}=\frac{1-\epsilon}{\epsilon},\tag{22}
\end{align*}
where $\tilde{\mathbf{G}}_{i,j}^{qp}\in \mathbb{C} ^{N^{\mathrm{dp}}\times M^{\mathrm{dp}}}$ denotes the channel coefficient matrix for the $i$-th subcarrier, from the TX-DPSIM polarization $p$ to the $j$-th user's RX-DPSIM polarization $q$. $0 < \epsilon \leq 1$ is the proportion of radiated power conversed from polarization 1 to 0 and vice versa. Therefore, the dual-polarized channel coefficient matrix $\mathbf{G}_{i,j}^{\mathrm{dp}}\in \mathbb{C} ^{N^{\mathrm{dp}}\times M^{\mathrm{dp}}}$ is modeled as ~\cite{9497725,10675359}
\begin{align*}
\label{eq23}
&\mathbf{G}_{i,j}^{\mathrm{dp}}=\left[ \begin{matrix}
	\mathbf{G}_{i,j}^{00}&		\mathbf{G}_{i,j}^{01}\\
	\mathbf{G}_{i,j}^{10}&		\mathbf{G}_{i,j}^{11}\\
\end{matrix} \right], \tag{23}
\\
\label{eq24}
&\mathbf{G}_{i,j}^{00}=e^{\mathrm{j}\psi ^{00}}\mathbf{G}_{i,j},\tag{24}
\\
\label{eq25}
&\mathbf{G}_{i,j}^{01}=\small{\frac{1}{\sqrt{\mathrm{XPD}}}}e^{\mathrm{j}\psi ^{01}}\mathbf{G}_{i,j},\tag{25}
\\
\label{eq26}
&\mathbf{G}_{i,j}^{10}=\small{\frac{1}{\sqrt{\mathrm{XPD}}}}e^{\mathrm{j}\psi ^{10}}\mathbf{G}_{i,j},\tag{26}
\\
\label{eq27}
&\mathbf{G}_{i,j}^{11}=e^{\mathrm{j}\psi ^{11}}\mathbf{G}_{i,j},\tag{27}
\end{align*}
where $\psi^{00}$, $\psi^{01}$, $\psi^{10}$, and $\psi^{11}$ respectively represent the phase shifts of the corresponding polarizations, $\mathbf{G}_{i,j}\in \mathbb{C} ^{N^{\mathrm{dp}}\times M^{\mathrm{dp}}}$ can be obtained from \eqref{eq9} in Section~\ref{II.~B}. 

%$\mathrm{XPD}\small{=\frac{1-\epsilon}{\varepsilon}}$$

\section{E2E problem-solving.} \label{Ⅲ}

\subsection{Problem posing.}  \label{Ⅲ.~A}
%我们定义多用户端到端系统的接收比特向量$\hat{\mathbf{b}}=\left[ \hat{\mathbf{b}}_1,\hat{\mathbf{b}}_2,\cdots ,\hat{\mathbf{b}}_J \right] ^{\mathrm{T}}$，我们的目标是尽可能的从接收向量$\hat{\mathbf{b}}$中准确恢复出发送向量$\mathbf{b}$。为此，SIM尝试主动调整超标面EM单元的相位来构造匹配信道，从而抑制由于物理信道中多径效应、频率选择性衰落等因素所造成的误码问题。 具体的，目标函数为
We define the received bit vector for a multi-user E2E system as $\hat{\mathbf{b}}=\left[ \hat{\mathbf{b}}_{1}^{\mathrm{T}},\hat{\mathbf{b}}_{2}^{\mathrm{T}},\cdots ,\hat{\mathbf{b}}_{J}^{\mathrm{T}} \right] ^{\mathrm{T}}$. Our goal is to recover the transmitted vector $\mathbf{b}$ from the received vector $\hat{\mathbf{b}}$ as accurately as possible. To this end, SIM (DPSIM) attempts to actively adjust the EM (DPEM) units on its metasurface to construct a matched channel, thereby suppressing bit error problems caused by factors such as multipath effects and frequency-selective fading. Specifically, the objective function is to minimize the bit error rate (BER) between $\mathbf{b}$ and $\hat{\mathbf{b}}$ , which is expressed as
\begin{align*}
\label{eq28}
\small{\mathrm{BER}\left( \mathbf{b},\hat{\mathbf{b}} \right) =\frac{1}{N^{\mathrm{bit}}}}\sum_{b=1}^{N^{\mathrm{bit}}}{P\left\{ \left[ \mathbf{b} \right] _b\ne \left[ \hat{\mathbf{b}} \right] _b \right\}}.\tag{28}
\end{align*}
Subsequently, the optimization problem for the SIM-assisted E2E OFDM system is formulated as follows
% 然后，SIM辅助的E2E OFDM系统的优化问题如下
\begin{align*}
\label{eq29a}
\mathcal{P} 1:&\min_{\mathbf{\Phi }^l,\mathbf{\Psi }_{j}^{k},\mathbf{p}_i,\mathbf{x}_i} \mathrm{BER}\left( \mathbf{b},\hat{\mathbf{b}} \right) ,\tag{29a}
\\
\label{eq29b}
\mathrm{s}.\mathrm{t}. &\sum_{i=1}^{N^{\mathrm{c}}}{\left\| \mathbf{p}_i \right\| _1}=P^{\mathrm{t}},\tag{29b}
\\
\label{eq29c}
&\mathbb{E} \left\{ \mathbf{xx}^{\mathrm{H}} \right\} =\mathbf{I}_{A^{\mathrm{t}}N^{\mathrm{c}}},\tag{29c}
\\
\label{eq29d}
&\left( 1-15 \right) ,\left( 28 \right),\tag{29d}
\end{align*}
%其中$P^{\mathrm{t}}$表示系统的发射总功率。同理，DPSIM辅助HMIMO系统的E2E OFDM系统的优化问题如下
where $P^{\mathrm{t}}$ denotes the total transmit power of the system. Similarly, the optimization problem for the DPSIM-assisted E2E OFDM system is formulated as follows
\begin{align*}
\label{eq30a}
\mathcal{P} 2:&\min_{\mathbf{\Phi }_{p}^{l},\mathbf{\Psi }_{j,p}^{k},\mathbf{p}_{i}^{\mathrm{dp}},\mathbf{x}_{i}^{\mathrm{dp}}} \mathrm{BER}\left( \mathbf{b},\hat{\mathbf{b}} \right) ,\tag{30a}
\\
\label{eq30b}
\mathrm{s}.\mathrm{t}. &\sum_{i=1}^{N^{\mathrm{c}}}{\left\| \mathbf{p}_{i}^{\mathrm{dp}} \right\| _1}=P^{\mathrm{t}},\tag{30b}
\\
\label{eq30c}
&\mathbb{E} \left\{ \mathbf{x}^{\mathrm{dp}}\left( \mathbf{x}^{\mathrm{dp}} \right) ^{\mathrm{H}} \right\} =\mathbf{I}_{2A^{\mathrm{dp},\mathrm{t}}N^{\mathrm{c}}},\tag{30c}
\\
\label{eq30d}
&\left( 3-4 \right) ,\left( 9-28 \right).\tag{30d}
\end{align*}
%由于优化变量之间的强耦合以及超表面EM(DPEM)单元透射系数的非凸单位模约束，获得问题$\mathcal{P} 1$和$\mathcal{P} 2$的全局最优解是极具挑战性的。有趣的是，我们发现该E2E系统与AE结构相似，因此我们提出可以使用深度学习方法来训练该系统，旨在实现灵活调制、解调的同时建立SIM (DPSIM)的自动调控机制。
Due to the strong coupling among the optimization variables and the non-convex unit-modulus constraints on the transmission coefficients of the metasurface EM (DPEM) units, obtaining the globally optimal solutions of problems $\mathcal{P}1$ and $\mathcal{P}2$ is extremely challenging. Interestingly, we observe that the considered E2E system exhibits a structural similarity to an AE. Motivated by this observation, we propose to exploit deep learning techniques to train the system, aiming to realize flexible modulation and demodulation, while simultaneously establishing an autonomous control mechanism for the SIM (DPSIM).

\subsection{Design of EMNN.} \label{Ⅲ.~B}

\begin{figure*}[!t]
\centering
\includegraphics[width=7in]{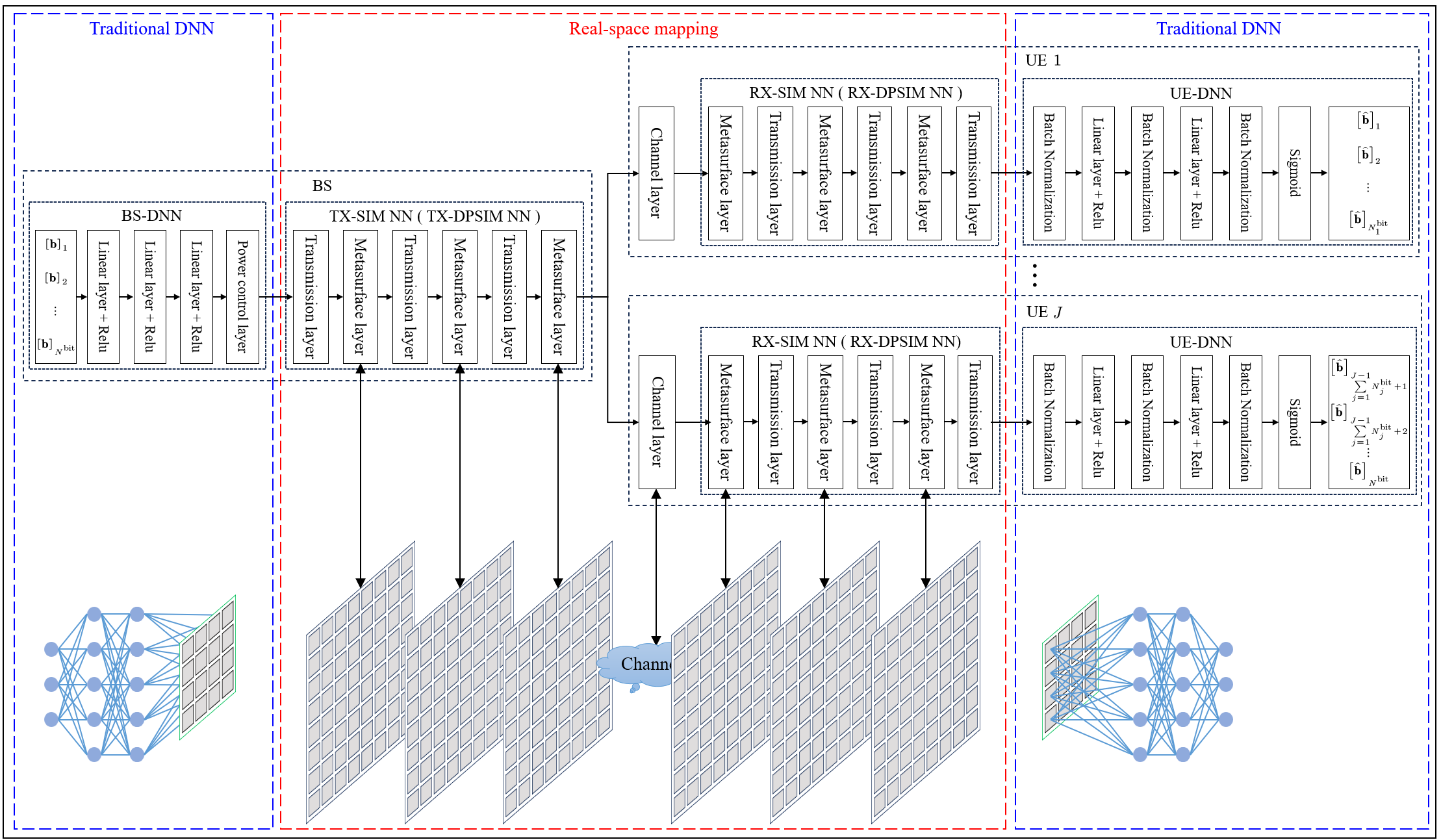}
\caption{Schematic diagram of the EMNN structure.}
\label{Fig6}
\vspace{-0.3cm}
\end{figure*}

%As shown in Fig. \ref{Fig6}，我们所提出的EMNN的核心思想是将SIM (DPSIM)中的超表面层抽象为若干可以用于训练的隐藏层，将真实的电磁传播抽象为深度学习网络里的前向传播过程。通过随机生成大量的发送比特向量$\mathbf{b}$，并采用mini-batch  Stochastic Gradient Descent (SGD) algorithm来训练EMNN。训练完成后，提取相应隐藏层的权重并对应映射至SIM (DPSIM)器件中，从而对每一个EM（DPEM）单元进行精确调控。综上，该方法整体上实现了artificial intelligence (AI)模型驱动控制的效果。除此之外，我们还在发送比特向量到发送信号之间，接收信号到接收比特向量之间各自添加了传统的DNN，以实现自动调制和解调。 
As shown in Fig.~\ref{Fig6}, the core idea of the proposed EMNN is to abstract the metasurface layer in the SIM (DPSIM) into several trainable hidden layers, while modeling the actual EM wave propagation as the forward propagation process in a deep learning network. A large number of transmit bit vectors $\mathbf{b}$ are randomly generated, and the mini-batch stochastic gradient descent (MBSGD) algorithm is employed to train the EMNN. Upon completion of the training, the weights of the corresponding hidden layers are extracted and mapped to the SIM (DPSIM) device, thereby enabling precise control of each EM (DPEM) unit. In summary, this approach essentially achieves an artificial intelligence (AI)-driven control mechanism. Furthermore, conventional DNNs are implemented between the transmit bit vectors and the transmitted signals, as well as between the received signals and the recovered bit vectors, to realize automatic modulation and demodulation.

%考虑到信号通常被刻画为复数形式，而神经网络计算一般为实值计算。因此，对于一根单极化天线（或双极化天线一个极化方向）的发射信号和接收信号，其实部和虚部各自对应隐藏层的一个权重；对于一个EM单元（或DPEM单元一个极化方向），由于其模长恒为1，因此仅需建立相位系数与隐藏层权重的双射关系。EMNN可以划分为四个子网络和一个信道运算层，我们依次详细介绍，EMNN的参数设计可以参考Table~\ref{table1}。
Considering that signals are typically represented in the complex domain, whereas neural network computations are generally performed in the real domain, the real and imaginary parts of the transmit and receive signals for a single-polarized antenna (or one polarization of a dual-polarized antenna) are mapped to two separate weights in the hidden layer. For an EM unit (or one polarization of a DPEM unit), since its magnitude is always unity, only a bijective mapping between its phase coefficient and the corresponding hidden-layer weight is required. The EMNN can be divided into four sub-networks and one channel computation layer, which will be introduced in detail in the following subsections. The parameter configuration of the EMNN is summarized in Table~\ref{table1}.

\textbf{Sub-network 1. BS-DNN.} 
%这是一个传统的DNN，其被直接部署在BS发射机中以实现发送比特向量$\mathbf{b}$到TX发射信号$\mathbf{x}\odot \mathbf{p}$或$\mathbf{x}^{\mathrm{dp}}\odot \mathbf{p}^{\mathrm{dp}}$的映射。其中
This is a conventional DNN, which is directly deployed at the BS transmitter to realize the mapping from the transmit bit vector $\mathbf{b}$ to the TX transmit signal $\mathbf{x}\odot \mathbf{p}$ or $\mathbf{x}^{\mathrm{dp}}\odot \mathbf{p}^{\mathrm{dp}}$, where
\begin{align*}
\label{eq31}
&\mathbf{p}=\left[ \left( \mathbf{p}_1 \right) ^{\mathrm{T}},\left( \mathbf{p}_2 \right) ^{\mathrm{T}},\cdots ,\left( \mathbf{p}_{N^{\mathrm{c}}} \right) ^{\mathrm{T}} \right] ^{\mathrm{T}},\tag{31}
\\
\label{eq32}
&\mathbf{p}^{\mathrm{dp}}=\left[ \left( \mathbf{p}_{1}^{\mathrm{dp}} \right) ^{\mathrm{T}},\left( \mathbf{p}_{2}^{\mathrm{dp}} \right) ^{\mathrm{T}},\cdots ,\left( \mathbf{p}_{N^{\mathrm{c}}}^{\mathrm{dp}} \right) ^{\mathrm{T}} \right] ^{\mathrm{T}}.\tag{32}
\end{align*}
%DNN中包含若干个线性隐藏层，并均使用Relu激活。最后，功率控制层是一个固定的运算层，其根据给定的发射总功率$P^{\mathrm{t}}$对信号进行按比例幅值缩放。
The DNN consists of several linear hidden layers, all employing the ReLU activation function. Moreover, the power control layer is a fixed computational layer that scales the signal amplitude proportionally according to the given total transmit power $P^{\mathrm{t}}$.

\textbf{Sub-network 2. TX-SIM (TX-DPSIM) NN.} 
%这是一个特殊的全连接神经网络，其目标是构建与TX-SIM/TX-DPSIM真实器件的双射关系。该网络由从系统模型中抽象得到的transmission layer和超表面层交替构成。Transmission layer可以由~\eqref{eq3}和~\eqref{eq4}直接计算得到，是不可训练的固定运算层。超表面层与TX-SIM (TX-DPSIM) 器件中的真实超表面一一对应，是可以用于训练的隐藏层。
This is a specialized fully-connected neural network designed to establish a bijective relationship with the actual TX-SIM (TX-DPSIM) hardware. The network is composed of alternating transmission layers and metasurface layers, both abstracted from the system model. The transmission layers can be directly computed from~\eqref{eq3} and~\eqref{eq4}, and are implemented as non-trainable fixed computation layers. In contrast, the metasurface layers correspond one-to-one with the physical metasurfaces in the TX-SIM (TX-DPSIM) device and are implemented as trainable hidden layers.

\textbf{Channel layer.} 
%信道层是一个通过信道估计、信道预测等手段从真实空间中获取的固定运算层。考虑到信道的时变特性，我们设计了统计信道和瞬时信道两种信道运算层，并使用迁移学习的方法来训练整个模型，具体训练步骤将在Section~\ref{Ⅲ.~C}中详细介绍。
The channel layer is a fixed computational layer obtained from the real world via techniques such as channel estimation or channel prediction. Considering the time-varying characteristics of the channel, we design two types of channel computation layers: a statistical channel and an instantaneous channel. Then, we employ transfer learning to train the entire model, and the specific training procedure will be detailed in Section~\ref{Ⅲ.~C}.

\textbf{Sub-network 3. RX-SIM (RX-DPSIM) NN.} 
%与TX-SIM (TX-DPSIM) NN的设计类似。这也是一个特殊的全连接神经网络，其目标是构建与RX-SIM (RX-DPSIM)真实器件的双射关系。不同的是，该子网路再次由$J$个模块构成，每个模块依次对应各个UE所集成的RX-SIM （RX-DPSIM）器件。transmission layer和超表面层的设计思想与前文相同，这里就不再介绍。
Similar to the design of the TX-SIM (TX-DPSIM) NN, this is also a specialized fully-connected neural network aimed at establishing a bijective relationship with the actual RX-SIM (RX-DPSIM) hardware. The difference lies in that this sub-network is further composed of $J$ modules, each corresponding to the RX-SIM (RX-DPSIM) device integrated at a specific UE. The design principles of the transmission layers and metasurface layers are identical to those described earlier and are therefore omitted here.

\begin{table}[!t] 
    \centering
    \caption{Output dimensions (OD) of each layer of EMNN(All the complex numbers are split into their real and imaginary parts for calculation).}
    \label{table1}
    \renewcommand{\arraystretch}{0.9} 
    \begin{tabular}{cccc} % {l|c|r} 定义了三列：
                           % l: 左对齐 (left-aligned)
                           % c: 居中对齐 (center-aligned)
                           % r: 右对齐 (right-aligned)
                           % |: 表示列之间有竖线
        \toprule
                \textbf{Module} & \textbf{Layer} & \textbf{OD (SIM)} & \textbf{OD (DPSIM)} \\ % 表头行，使用 & 分隔列，\\ 结束行
        \midrule
                & Input & $N^{\mathrm{bit}}$ & $N^{\mathrm{bit}}$\\
          & Liner layer + Relu & $N^{\mathrm{bit}}N^{\mathrm{c}}$ & $N^{\mathrm{bit}}N^{\mathrm{c}}$\\
         BS-DNN       & Liner layer + Relu & $N^{\mathrm{bit}}N^{\mathrm{c}}$ & $N^{\mathrm{bit}}N^{\mathrm{c}}$\\
                & Liner layer + Relu & $2A^{\mathrm{t}}N^{\mathrm{c}}$ & $4A^{\mathrm{dp,t}}N^{\mathrm{c}}$\\
                & Power control layer & $2A^{\mathrm{t}}N^{\mathrm{c}}$ & $4A^{\mathrm{dp,t}}N^{\mathrm{c}}$\\
        \midrule
                & Transmission layer $1$ & $2M$ & $4M^{\mathrm{dp}}$\\
                & Metasurface layer $1$ & $2M$ & $4M^{\mathrm{dp}}$\\
                & Transmission layer $2$ & $2M$ & $4M^{\mathrm{dp}}$\\
TX-SIM          & Metasurface layer $2$ & $2M$ & $4M^{\mathrm{dp}}$\\
(TX-DPSIM)       & ... & ... & ...\\
NN                & ... & ... & ...\\
                & Transmission layer $L$ & $2M$ & $4M^{\mathrm{dp}}$\\
                & Metasurface layer $L$ & $2M$ & $4M^{\mathrm{dp}}$\\
        \midrule
        Channel & Channel layer & $2JN$ & $4JN^{\mathrm{dp}}$\\   
        \midrule
                & Metasurface layer $K$ &$2JN$ & $4JN^{\mathrm{dp}}$\\
                & Transmission layer $K$ &$2JN$ & $4JN^{\mathrm{dp}}$\\
                & Metasurface layer $K-1$ &$2JN$ & $4JN^{\mathrm{dp}}$\\
RX-SIM          & Transmission layer $K-1$ &$2JN$ & $4JN^{\mathrm{dp}}$\\
(RX-DPSIM)       & ... & ... & ...\\
NN                & ... & ... & ...\\
                & Metasurface  layer $1$ & $2JN$ & $4JN^{\mathrm{dp}}$\\
                & Transmission layer $1$ & $2JA^{\mathrm{r}}N^{\mathrm{c}}$ & $4JA^{\mathrm{dp,r}}N^{\mathrm{c}}$\\
        \midrule
                & Batch Normalization & $2JA^{\mathrm{r}}N^{\mathrm{c}}$ & $4JA^{\mathrm{dp,r}}N^{\mathrm{c}}$\\
                & Liner layer + Relu & $N^{\mathrm{bit}}N^{\mathrm{c}}$ & $N^{\mathrm{bit}}N^{\mathrm{c}}$\\
          & Batch Normalization & $N^{\mathrm{bit}}N^{\mathrm{c}}$ & $N^{\mathrm{bit}}N^{\mathrm{c}}$\\
         UE-DNN       & Liner layer + Relu & $N^{\mathrm{bit}}$ & $N^{\mathrm{bit}}$\\
                & Batch Normalization & $N^{\mathrm{bit}}$ & $N^{\mathrm{bit}}$\\
                & Sigmoid & $N^{\mathrm{bit}}$ & $N^{\mathrm{bit}}$\\
                & Output & $N^{\mathrm{bit}}$ & $N^{\mathrm{bit}}$\\
        \bottomrule
    \end{tabular}
\end{table}

\begin{figure*}[b]
\vspace{-0.7cm}
\begin{align*}
&\overline{\ \ \ \ \ \ \ \ \ \ \ \ \ \ \ \ \ \ \ \ \ \ \ \ \ \ \ \ \ \ \ \ \ \ \ \ \ \ \ \ \ \ \ \ \ \ \ \ \ \ \ \ \ \ \ \ \ \ \ \ \ \ \ \ \ \ \ \ \ \ \ \ \ \ \ \ \ \ \ \ \ \ \ \ \ \ \ \ \ \ \ \ \ \ \ \ \ \ \ \ \ \ \ \ \ \ \ \ \ \ \ \ \ \ \ \ \ \ \ \ \ \ \ \ \ \ \ \ \ \ \ \ \ \ \ \ \ \ \ \ \ \ \ \ \ \ \ }\\
\label{eq31}
&\ \ \ \ \ \ \ \ \ \ \ \ \ \ \ \ \ \ \mathcal{L} \left( \mathbf{b},\hat{\mathbf{b}} \right) =-\frac{1}{N^{\mathrm{batch}}}\sum_{\mathfrak{b} =1}^{N^{\mathrm{batch}}}{\sum_{b=1}^{N^{\mathrm{bit}}}{\left\{ \left[ \mathbf{b}_{\mathfrak{b}} \right] _b\log \left( \left[ \hat{\mathbf{b}}_{\mathfrak{b}} \right] _b \right) +(1-\left[ \mathbf{b}_{\mathfrak{b}} \right] _b)\log \left( 1-\left[ \hat{\mathbf{b}}_{\mathfrak{b}} \right] _b \right) \right\}}}.\tag{31}
\end{align*}
\end{figure*}

\textbf{Sub-network 4. UE-DNN.} 
%与BS-DNN类似，这也是一个传统DNN。不同的是，该子网路再次由$J$个模块构成，每个模块被依次部署在各个UE中以实现RX接收信号$\mathbf{y}_j$或$\mathbf{y}_{j}^{\mathrm{dp}}$到接收比特向量$\hat{\mathbf{b}}_j$的映射。DNN交替执行批次归一化和线性层运算，线性层均使用Relu激活。最后，使用Sigmoid函数将输出归一化至$\left[ 0,1 \right]$,并使用硬判决恢复比特向量。
Similar to the BS-DNN, this is also a conventional DNN. The difference is that this sub-network is further divided into $J$ modules, each deployed at an individual UE to realize the mapping from the received signal $\mathbf{y}_j$ or $\mathbf{y}_{j}^{\mathrm{dp}}$ to the detected bit vector $\hat{\mathbf{b}}_j$. The DNN alternates between batch normalization and linear layers, with all linear layers employing the ReLU activation function. Finally, a sigmoid function is applied to normalize the output to the range $\left(0,1\right)$, followed by hard decision to recover the bit vector.

\subsection{Model training and deployment framework.} \label{Ⅲ.~C}
%为了在时变信道中准确恢复原始比特向量，E2E系统需要每隔一段时间根据CSI信息进行重复训练。对此，我们可以采用迁移学习的方法来加快训练速度。具体的，首先通过统计CSI来训练基础模型，然后根据瞬时CSI来对模型进行实时微调。我们在模型训练过程中使用的损失函数为二元交叉熵（BCE），可以通过~\eqref{eq31}计算得到，其中$N^{\mathrm{batch}}$表示训练的batch size。与\eqref{eq28}类似，BCE用于估计输入与输出的一致性。$\mathcal{L} (\mathbf{b},\hat{\mathbf{b}})\geqslant 0$当且仅当$\mathbf{b}=\hat{\mathbf{b}}$等号成立。因此BCE越小，比特流恢复越准确，系统性能更好。
For accurately recovering the original bit vector in time-varying channels, the E2E system needs to be retrained periodically based on CSI information. To accelerate this process, we can employ transfer learning. Specifically, a base model is first trained using the statistical CSI, and then the model is fine-tuned in real time according to the instantaneous CSI. The loss function used during the model training is the binary cross-entropy (BCE), which can be computed as in~\eqref{eq31}, where $N^{\mathrm{batch}}$ denotes the training batch size. Similar to~\eqref{eq28}, BCE is employed to evaluate the consistency between the input and the output. It holds that $\mathcal{L} (\mathbf{b},\hat{\mathbf{b}})\geqslant 0$ with equality if and only if $\mathbf{b}=\hat{\mathbf{b}}$. Therefore, a smaller BCE indicates more accurate bitstream recovery and thus better system performance.

%我们通过MBSGD算法在线更新EMNN各子网络的权重，BS-DNN、TX-SIM NN、RX-SIM NN和UE-DNN的权重更新公式依次如下
We update the weights of each sub-network in the EMNN online via the MBSGD algorithm, 
with the weight update formulas for BS-DNN, TX-SIM NN, RX-SIM NN, and UE-DNN given respectively as
\begin{align*}
\label{eq32}
&\mathbf{W}^{\mathrm{T}}\gets \mathbf{W}^{\mathrm{T}}-\eta \nabla _{\mathbf{W}^{\mathrm{T}}}\mathcal{L} \left( \mathbf{b},\hat{\mathbf{b}} \right),\tag{32}
\\
\label{eq33}
&\mathbf{\Theta }^l\gets \mathbf{\Theta }^l-\eta \nabla _{\mathbf{\Theta }^l}\mathcal{L} \left( \mathbf{b},\hat{\mathbf{b}} \right) ,1\leqslant l\leqslant L,\tag{33} 
\\
\label{eq34}
&\mathbf{\Xi }^k\gets \mathbf{\Xi }^k-\eta \nabla _{\mathbf{\Xi }^k}\mathcal{L} \left( \mathbf{b},\hat{\mathbf{b}} \right) ,1\leqslant k\leqslant K,\tag{34}
\\
\label{eq35}
&\mathbf{W}^{\mathrm{R}}\gets \mathbf{W}^{\mathrm{R}}-\eta \nabla _{\mathbf{W}^{\mathrm{R}}}\mathcal{L} \left( \mathbf{b},\hat{\mathbf{b}} \right),\tag{35}
\end{align*}
%其中$\eta$表示学习率，$\mathbf{W}^{\mathrm{T}}$和$\mathbf{W}^{\mathrm{R}}$分别表示BS-DNN和UE-DNN的权重张量。$\mathbf{\Theta }^l$和$\mathbf{\Xi }^k$分别为TX-SIM和RX-SIM中超表面抽象后对应隐藏层的权重张量。类似地，对于DPSIM辅助的系统，\eqref{eq33}和\eqref{eq34}对应的更新公式为
where $\eta$ denotes the learning rate, $\mathbf{W}^{\mathrm{T}}$ and $\mathbf{W}^{\mathrm{R}}$ 
respectively represent the weight tensors of the BS-DNN and UE-DNN, $\mathbf{\Theta }^l$ and $\mathbf{\Xi }^k$ are the weight tensors of the corresponding hidden layers abstracted from the metasurfaces in TX-SIM and RX-SIM, respectively. Similarly, for the DPSIM-assisted system, the update formulas corresponding to \eqref{eq33} and \eqref{eq34} are given by
\begin{align*}
\label{eq36}
&\mathbf{\Theta }_{p}^{l}\gets \mathbf{\Theta }_{p}^{l}-\eta \nabla _{\mathbf{\Theta }_{p}^{l}}\mathcal{L} \left( \mathbf{b},\hat{\mathbf{b}} \right),1\leqslant l\leqslant L^{\mathrm{dp}},\tag{36}
\\
\label{eq37}
&\mathbf{\Xi }_{p}^{k}\gets \mathbf{\Xi }_{p}^{k}-\eta \nabla _{\mathbf{\Xi }_{p}^{k}}\mathcal{L} \left( \mathbf{b},\hat{\mathbf{b}} \right),1\leqslant k\leqslant K^{\mathrm{dp}},\tag{37}
\end{align*}
%其中$\mathbf{\Theta }_{p}^{l}$和$\mathbf{\Xi }_{p}^{k}$分别为TX-DPSIM和RX-DPSIM中超表面不同极化方向抽象后对应隐藏层的权重张量。
where $\mathbf{\Theta }_{p}^{l}$ and $\mathbf{\Xi }_{p}^{k}$ are the weight tensors of the hidden layers corresponding to the different polarization directions of the metasurface in the TX-DPSIM and RX-DPSIM after abstraction, respectively.

\begin{figure}[!t]
\centering
\includegraphics[width=3.5in]{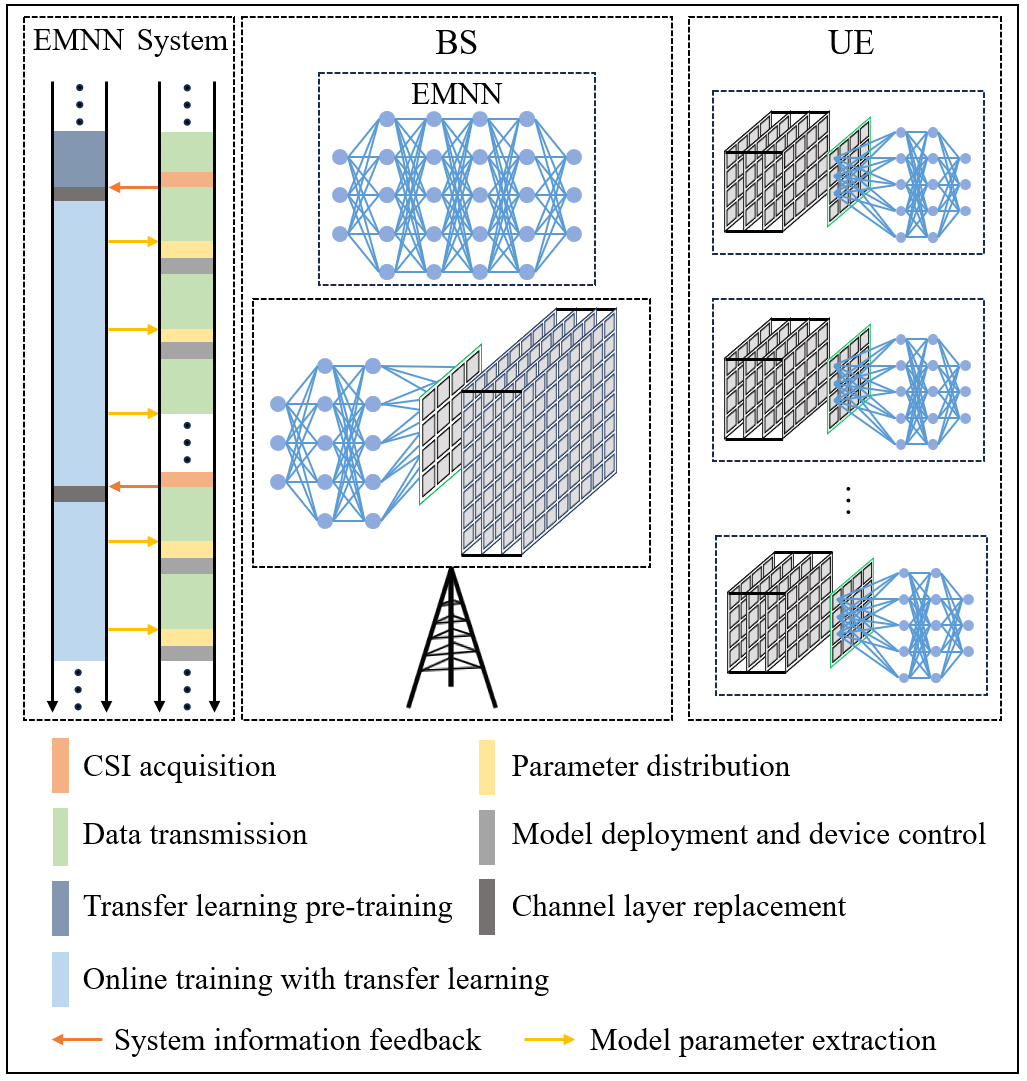}
\caption{Model training and deployment framework.}
\label{Fig7}
\end{figure}

%As shown in Fig. \ref{Fig7}，我们提出了EMNN的一个训练和部署框架。重要的是，EMNN将会始终在BS侧训练。每隔一段时间，BS会将RX-SIM (RX-DPSIM) NN和UE-DNN的权重剪枝量化后各自分发给对应用户。最后，BS发射机实时部署BS-DNN并根据TX-SIM (TX-DPSIM) NN的权重实时调控EM (DPEM)单元。各个用户则根据接收到的权重部署UE-DNN并同时动态调控RX-SIM (RX-DPSIM)中的EM (DPEM)单元。EMNN的训练和部署步骤具体为
As shown in Fig.~\ref{Fig7}, we propose a training and deployment framework for EMNN. Importantly, EMNN is always trained on the BS side. At regular intervals, the BS prunes and quantizes the weights of the RX-SIM (RX-DPSIM) NN and the UE-DNN, and then distributes them to the corresponding users. Finally, the BS transmitter deploys the BS-DNN in real time and dynamically controls the EM (DPEM) units according to the weights of the TX-SIM (TX-DPSIM) NN. Each user, upon receiving the weights, deploys the UE-DNN and simultaneously adjusts the EM (DPEM) units in the RX-SIM (RX-DPSIM) dynamically. The specific training and deployment procedures of EMNN are as follows

\textbf{Step 1. Pre-training of the EMNN.} 
%在一个长时间周期开始，我们假设BS已知所有服务用户的统计CSI。因此在实时传输数据之前，EMNN通过统计信道层来进行预训练以获得基础模型。
At the beginning of a long time period, we assume that the BS knows the statistical CSI of all service users. Therefore, before transmitting real-time data, the EMNN is pre-trained through the statistical channel layer to obtain the base model.

\textbf{Step 2. Instantaneous CSI acquisition and online training of EMNN.}
%BS通过信道估计、信道预测等手段获取各个用户的瞬时CSI。然后，EMNN通过瞬时信道层进行实时在线训练。 
The BS obtains the instantaneous CSI of each user through channel estimation or channel prediction. Then, the EMNN performs real-time online training through the instantaneous channel layer.

\textbf{Step 3. Parameter distribution.}
%每间隔一小段时间，BS将EMNN的权重进行提取和分割，将发射端对应的权重数据保留，而接收端对应权重数据对应分发给各个用户。
At short regular intervals, the BS extracts and partitions the EMNN weights, retaining the weights corresponding to the transmitter side, while distributing the receiver-side weights to the respective users.

\textbf{Step 4. Systems control.}
%BS-DNN和UE-DNN是传统神经网络，因此可以根据分发得到的权重数据直接在收发机内部署。TX-SIM（TX-DPSIM）和RX-SIM (RX-DPSIM)器件则需根据预定义的映射规则将权重数据逆变换，方可实现EM（DPEM）单元的实时调控。
BS-DNN and UE-DNN are conventional neural networks that can be directly deployed inside the transceivers based on the distributed weight data, whereas the TX-SIM (TX-DPSIM) and RX-SIM (RX-DPSIM) devices must apply an inverse transformation to the weight data according to predefined mapping rules before real-time control of the EM (DPEM) units.

\subsection{Complexity analysis.} \label{Ⅲ.~D}

%所提出EMNN训练的总体复杂度源于超表面网络（SIM-NN/DPSIM-NN）波域信号处理与传统神经网络（BS-DNN/UE-DNN）数字化计算的叠加。对于SIM辅助系统，其复杂度可分解为多个部分。BS-DNN与UE-DNN的复杂度取决于其网络结构，若其隐藏层的典型维度为$D$，则参数量约为$\mathcal{O}(D^2 H J)$，其中$H$为隐藏层数，$J$为用户数。其前向传播的计算复杂度主要为矩阵乘法，约为$\mathcal{O}(D^2 H J)$。超表面部分的参数量为TX-SIM的$\mathcal{O}(M L)$与所有用户RX-SIM的$\mathcal{O}(N K J)$。其计算复杂度则主要由模拟层间电磁传输的矩阵乘法主导，对于每个子载波，TX-SIM的复杂度约为$\mathcal{O}(M^2 L)$，所有RX-SIM的复杂度约为$\mathcal{O}(N^2 K J)$。因此，整个SIM辅助系统的总参数量为$\mathcal{O}(M L + N K J + D^2 H J)$，而单次前向传播的总体计算复杂度随子载波数$N^c$缩放，约为$\mathcal{O}(N^c (M^2 L + N^2 K J) + D^2 H J)$。类似的，对于DPSIM辅助系统，其架构可视为两个并行的SIM，因此大部分结构的复杂度约为SIM辅助系统的两倍。如果我们假设DPSIM辅助系统中DNN的网络结构与SIM辅助系统一致，则DPSIM辅助系统的总参数量约为$\mathcal{O}(2(M^{dp} L^{dp} + N^{dp} K^{dp} J + D^2 H J))$，总计算复杂度约为$\mathcal{O}(2( N^c ( (M^{dp})^2 L^{dp} + (N^{dp})^2 K^{dp} J ) + D^2 H J ))$。
The overall training complexity of the proposed EMNN arises from the superposition of wave-domain signal processing in the metasurface network and digital computation in conventional neural networks. For the SIM-assisted system, the overall complexity can be decomposed into several components. In particular, the complexities of the BS-DNN and UE-DNN are primarily determined by their respective network architectures. If the typical hidden layer dimension is denoted by $D$, the number of parameters is on the order of $\mathcal{O}(D^2 H J)$, where $H$ denotes the number of hidden layers. The computational complexity of forward propagation is mainly dominated by matrix multiplications and is also on the order of $\mathcal{O}(D^2 H J)$. For the metasurface part, the number of parameters is $\mathcal{O}(M L)$ for the TX-SIM and $\mathcal{O}(N K J)$ for all RX-SIMs at the users. The corresponding computational complexity is dominated by matrix multiplications simulating the EM transmission between analog layers. For each subcarrier, the complexity of the TX-SIM is on the order of $\mathcal{O}(M^2 L)$, while that of all RX-SIMs is on the order of $\mathcal{O}(N^2 K J)$. Consequently, the total number of parameters of the SIM-assisted system is $\mathcal{O}(M L + N K J + D^2 H J)$, and the overall computational complexity of a single forward propagation, scaling with the number of subcarriers $N^{\mathrm{c}}$, is given by $\mathcal{O}(N^{\mathrm{c}} (M^2 L + N^2 K J) + D^2 H J)$. 

Similarly, for the DPSIM-assisted system, its architecture can be regarded as two parallel SIMs, resulting in most of its structural complexities being approximately twice those of the SIM-assisted system. Assuming that the DNN architectures in the DPSIM-assisted system are identical to those in the SIM-assisted system, the total number of parameters for the DPSIM-assisted system is $\mathcal{O}\!\left(2(M^{\mathrm{dp}} L^{\mathrm{dp}} + N^{\mathrm{dp}} K^{\mathrm{dp}} J) + D^2 H J\right)$, and the overall computational complexity is approximately $\mathcal{O} \!\left( 2N^{\mathrm{c}}(M^{\mathrm{dp}})^2L^{\mathrm{dp}}+2N^{\mathrm{c}}(N^{\mathrm{dp}})^2K^{\mathrm{dp}}J+D^2HJ \right)$.

\subsection{Model scalability.} \label{Ⅲ.~E}
%\textbf{Propagation Coefficient.} 由于实际硬件不可避免的制造缺陷和建模误差，Section~\ref{II.~A} 中相邻超表面的传播系数可能会偏离\eqref{eq3}和\eqref{eq4}。因此在集成SIM (DPSIM)器件之前，我们可以通过发送激励信号同时测量接收面板响应的方式预先矫正器件的传输系数矩阵~\cite{10158690}。
\textbf{Propagation coefficient.} Due to inevitable manufacturing imperfections and modeling errors in practical hardware, the propagation coefficients between adjacent metasurfaces in Section~\ref{II.~A} may deviate from \eqref{eq3} and \eqref{eq4}. Therefore, before integrating the SIM (DPSIM) device, we can pre-calibrate the transmission coefficient matrix of the device by transmitting excitation signals and simultaneously measuring the response of the receiving panel~\cite{10158690}.

%\textbf{EMNN structure.} 为实现优异的系统性能，当UE、子载波、或单次传输的比特数量变多时，BS-DNN和UE-DNN通常会需要更庞大的网络结构。我们设计的EMNN结构实现了子网路之间的解耦合，即可以将BS-DNN和UE-DNN直接替换为其它高性能网络。具体应该选择什么样的网络超出了本文的研究范围，留待后续研究。
\textbf{EMNN structure.} To achieve superior system performance, larger network architectures are typically required for the BS-DNN and UE-DNN as the number of UEs, subcarriers, or bits per transmission increases. The proposed EMNN architecture enables decoupling between the subnetworks, allowing the BS-DNN and UE-DNN to be directly replaced with other high-performance networks. The specific choice of such networks is beyond the scope of this work and is left for future research.

%\textbf{SIM (DPSIM) device.} 为便于读者理解，在前文中我们假设所有UE所集成的RX-SIM (RX-DPSIM)器件是一致的，且单个器件由完全相同的超表面依次排列构成。事实上，EMNN中的超表面层可以由不同尺度的超表面抽象得到，各个UE所集成的器件也无需一致，仅需建立好EM（DPEM）单元与隐藏层权重的对应关系即可。
\textbf{SIM (DPSIM) device.} For ease of understanding, we have assumed in the preceding discussion that all UEs are equipped with identical RX-SIM (RX-DPSIM) devices, each composed of a sequence of completely identical metasurfaces. In practice, the metasurface layers in the EMNN can be abstracted from metasurfaces of different scales, and the devices integrated into different UEs do not need to be identical, as long as the correspondence between the EM (DPEM) units and the hidden-layer weights is properly established.

\section{Simulation Results.} \label{Ⅳ}

\subsection{Simulation settings.}  \label{Ⅳ.~A}

\begin{table}[!t] 
    \centering
    \caption{Simulation parameter settings}
    \label{table2}
    \renewcommand{\arraystretch}{0.9} 
    \begin{tabular}{ll} % {l|c|r} 定义了三列：
                           % l: 左对齐 (left-aligned)
                           % c: 居中对齐 (center-aligned)
                           % r: 右对齐 (right-aligned)
                           % |: 表示列之间有竖线
        \toprule
                \textbf{System parameters} & \textbf{Value} \\
        \midrule
                Center frequency ($f_0$) & 28 GHz \\
                Center frequency wavelength ($\lambda_0$) & 10.7 mm \\
                Bandwidth ($B$) & 100 MHz\\
                Number of subcarriers ($N^{\mathrm{c}}$) & 32\\
                Number of UE ($J$) & 3\\
                BS coordinates & (0, 0, 0) m\\
                UE1 coordinates & (10, 0, 20) m\\
                UE2 coordinates & (20, 0, 20) m\\
                UE3 coordinates & (0, 0, 30) m\\
                Number of bits for each UE in an OFDM symbol & [32, 16, 8]\\
                Monte carlo & 100 \\
        \toprule
                \textbf{Channel parameters} & \textbf{Value} \\
        \midrule
                Polarization conversion power ratio ($\epsilon$) & 0.2\\
                Cross-polarization discrimination (XPD) & 4\\
                Number of scatterers ($S$) & 100\\
                Path loss reference distance ($d_0$) & 1 m \\
                Path loss exponent ($b$) & 3.5 \\
                Path loss shadowing fading variance ($\delta$) & 9 dB \\
                Receiver noise ($\sigma^2$) & -110 dBm \\
                Random generation method of scatterers  & ~\cite{9541182}\\
        \toprule
                \textbf{SIM parameters} & \textbf{Value} \\
        \midrule
                Number of layers of TX-SIM ($L$) & 3 \\
                Number of layers of RX-SIM ($K$) & 3 \\
                Number of EM units per layer of TX-SIM \\ \ \ \ \ \ \ \ \ \ \ \ \ \ \ \ \ \ \ \ ($M=M^{\mathrm{x}}\times M^{\mathrm{y}}$) & $100=10\times 10$ \\
                Number of EM units per layer of RX-SIM \\ \ \ \ \ \ \ \ \ \ \ \ \ \ \ \ \ \ \ \ 
                ($N=N^{\mathrm{x}}\times N^{\mathrm{y}}$) & $100=10\times 10$ \\
                Number of TX antennas\\ \ \ \ \ \ \ \ \ \ \ \ \ \ \ \ \ \ \ \ ($A^{\mathrm{t}}=A^{\mathrm{t}^{\mathrm{x}}}\times A^{\mathrm{t}^{\mathrm{y}}}$) & $16=4\times 4$ \\
                Number of RX antennas per use \\ \ \ \ \ \ \ \ \ \ \ \ \ \ \ \ \ \ \ \ 
                ($A^{\mathrm{r}}=A^{\mathrm{r}^{\mathrm{x}}}\times A^{\mathrm{r}^{\mathrm{y}}}$) & $9=3\times 3$ \\
                Spacing of EM units in TX-SIM ($d^{\mathrm{t}}$) & $\lambda_0/2$ \\
                Spacing of EM units in RX-SIM ($d^{\mathrm{r}}$) & $\lambda_0/2$ \\
                TX-SIM layer spacing ($r^{\mathrm{t}}$) & $\lambda_0/2$ \\
                RX-SIM layer spacing ($r^{\mathrm{r}}$) & $\lambda_0/2$ \\
        \toprule
                \textbf{DPSIM parameters} & \textbf{Value} \\
        \midrule
                Number of layers of TX-DPSIM ($L^{\mathrm{dp}}$) & 3 \\
                Number of layers of RX-DPSIM ($K^{\mathrm{dp}}$) & 3 \\
                Number of DPEM units per layer of TX-DPSIM \\ \ \ \ \ \ \ \ \ \ \ \ \ \ \ \ \ \ \ \ ($M^{\mathrm{dp}}=M^{\mathrm{dp},\mathrm{x}}\times M^{\mathrm{dp},\mathrm{y}}$) & $100=10\times 10$ \\
                Number of DPEM units per layer of RX-DPSIM \\ \ \ \ \ \ \ \ \ \ \ \ \ \ \ \ \ \ \ \ 
                ($N^{\mathrm{dp}}=N^{\mathrm{dp},\mathrm{x}}\times N^{\mathrm{dp},\mathrm{y}}$) & $100=10\times 10$ \\
                Number of TX dual-polarized antennas\\ \ \ \ \ \ \ \ \ \ \ \ \ \ \ \ \ \ \ \ ($A^{\mathrm{dp},\mathrm{t}}=A^{\mathrm{dp},\mathrm{t}^{\mathrm{x}}}\times A^{\mathrm{dp},\mathrm{t}^{\mathrm{y}}}$) & $9=3\times 3$ \\
                Number of RX dual-polarized antennas per use \\ \ \ \ \ \ \ \ \ \ \ \ \ \ \ \ \ \ \ \ 
                ($A^{\mathrm{dp},\mathrm{r}}=A^{\mathrm{dp},\mathrm{r}^{\mathrm{x}}}\times A^{\mathrm{dp},\mathrm{r}^{\mathrm{y}}}$) & $4=2\times 2$ \\
                Spacing of DPEM units in TX-DPSIM ($d^{\mathrm{dp,t}}$) & $\lambda_0/2$ \\
                Spacing of DPEM units in RX-DPSIM ($d^{\mathrm{dp,r}}$) & $\lambda_0/2$ \\
                TX-DPSIM layer spacing ($r^{\mathrm{dp,t}}$) & $\lambda_0/2$ \\
                RX-DPSIM layer spacing ($r^{\mathrm{dp,r}}$) & $\lambda_0/2$ \\
        \toprule
                \textbf{Training and testing parameters} & \textbf{Value}\\ % 表头行，使用 & 分隔列，\\ 结束行
        \midrule
                Loss function & BCE\\
                Initialization & Xavier\\
                Optimizer & AdamW\\
                Training epoch ($E$)  & 2000\\
                Learning rate & $0.005$\\
                Batch size ($N^{\mathrm{batch}}$) & 1000\\
                Performance metric & BER\\
                Test scale & 100000\\
                Learning rate decay & 1 / 1.05\\
        \bottomrule
        
    \end{tabular}
\end{table}

%考虑到Fig.~\ref{Fig7}中所提框架需要复杂的跨层建模和系统级实现，我们在本节中重点验证其核心部分——基于迁移学习训练的EMNN方法在SIM辅助E2E OFDM系统和DPSIM辅助E2E OFDM系统中的有效性。该方法的性能对整体框架的有效性起决定性作用，因此其验证结果在很大程度上反映所提框架的可行性。具体的，我们假设信道为准静态块衰落，首先EMNN使用统计信道来训练基础模型。然后随机生成若干组瞬时信道，EMNN分别对应替换信道层并通过迁移学习的方式来对模型微调。最后，取不同信道条件下的测试结果均值作为系统性能指标。
Considering that the framework proposed in Fig.~\ref{Fig7} requires complex cross-layer modeling and system-level implementation, in this section we focus on validating its core component, the EMNN method trained via transfer learning in SIM-assisted E2E OFDM systems and DPSIM-assisted E2E OFDM systems. The performance of this method plays a decisive role in the overall effectiveness of the framework; therefore, its validation results largely reflect the feasibility of the proposed framework. Specifically, we assume a quasi-static block-fading channel. First, the EMNN uses statistical CSI to train a base model. Then, multiple sets of instantaneous channels are randomly generated, and for each set, the EMNN replaces the channel layer accordingly and fine-tunes the model via transfer learning. Finally, the average of the test results under different channel conditions is taken as the system performance metric.

\begin{figure}[!t]
\centering
\includegraphics[width=2.85in]{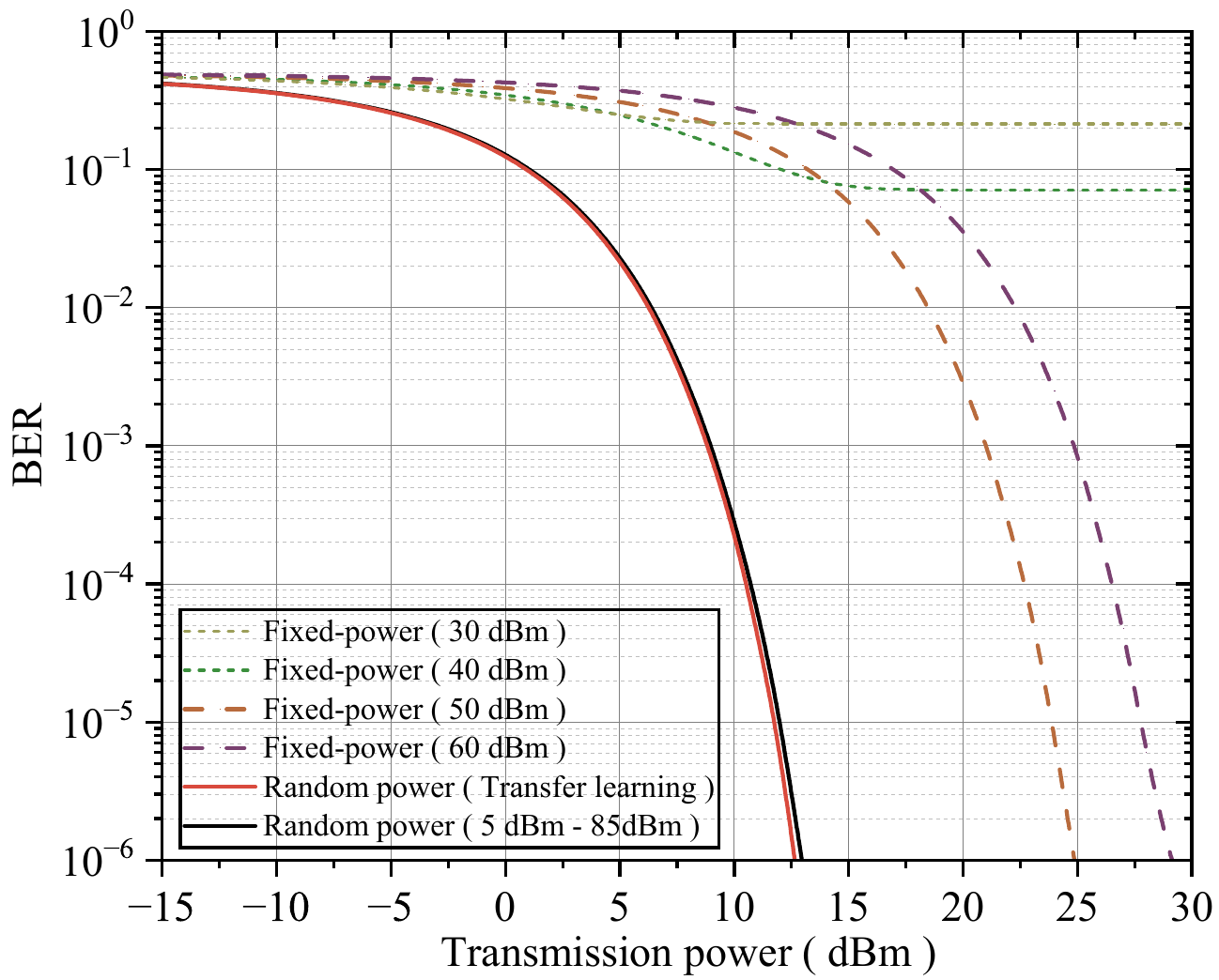}
\caption{Performance comparison of different training methods.}
\label{Fig8}
\vspace{-0.1cm}
\end{figure}

\begin{figure}[!t]
\centering
\includegraphics[width=2.85in]{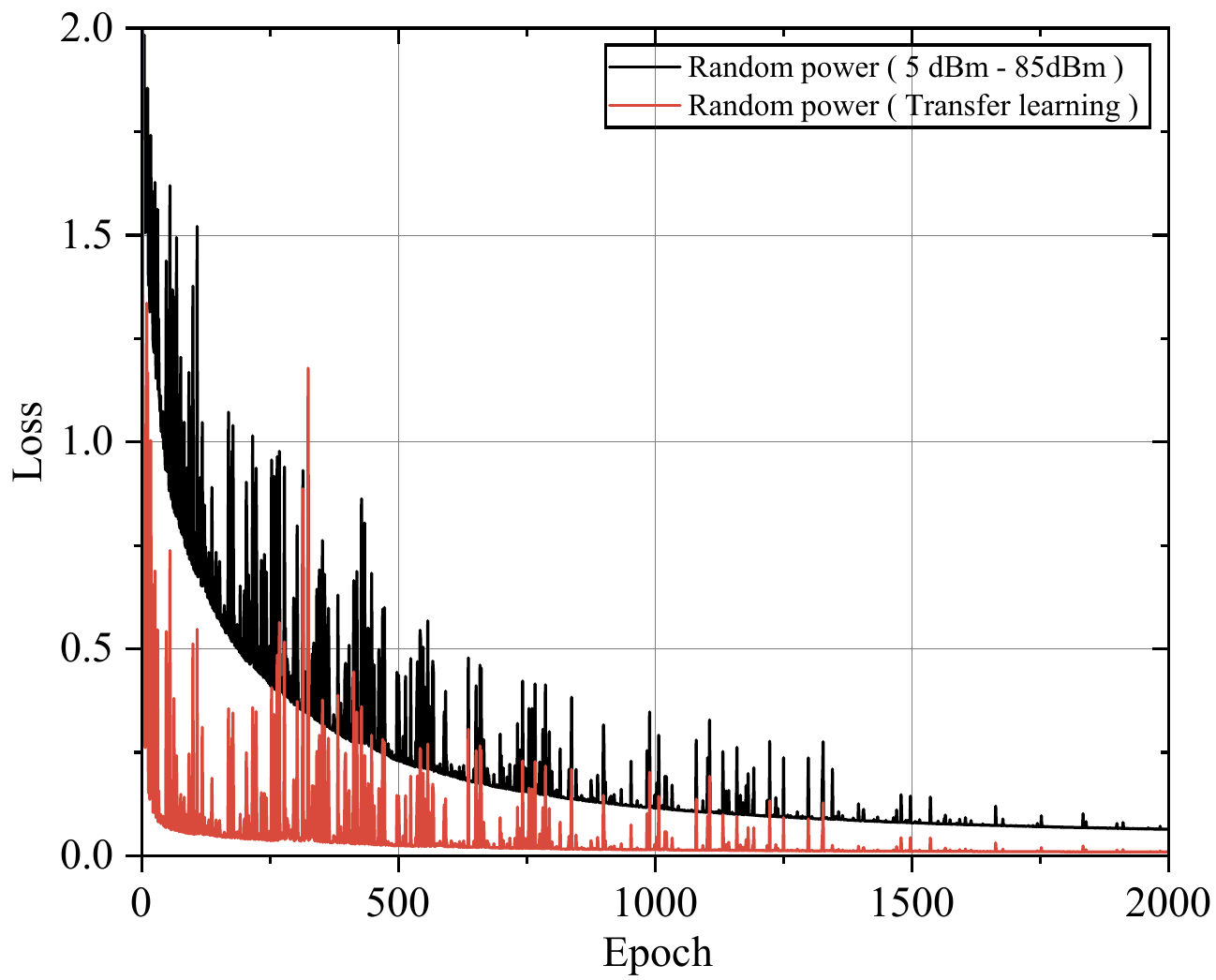}
\caption{Iteration graph of EMNN training loss.}
\label{Fig9}
\end{figure}

%\ref{table2}中依次给出了用于仿真验证的系统参数设置、信道参数设置、SIM (DPSIM)参数设置、模型训练和测试参数设置。后文的仿真结果除有特殊说明，均基于以上设置得到。
The path loss between the transmitter and the receiver is modeled by~\cite{7109864}
\begin{align*}
\label{eq38}
\text{PL}(d) = \text{PL}(d_0) + 10b \log_{10} \left(\frac{d}{d_0}\right) + X_\delta,\ d \geq d_0,\tag{38}
\end{align*}
where $\text{PL}(d_0) = 20 \log_{10} \left(\frac{4\pi d_0}{\lambda}\right) \text{ dB}$ is the free space path loss at the reference distance $d_0$, $b$ represents the path loss exponent, $X_\delta$ is a zero mean Gaussian random variable with a standard deviation $\delta$, characterizing the large-scale signal fluctuations of shadow fading. Furthermore, Table~\ref{table2} sequentially summarizes the system parameter settings, channel parameter settings, SIM (DPSIM) parameter settings, and model training (testing) parameter settings used for simulation. Unless otherwise specified, the subsequent simulation results are obtained based on the above settings.

\subsection{Model Training Comparative Analysis.} \label{Ⅳ.~B}

%Fig.~\ref{Fig8}展示了使用不同训练方法来对EMNN进行训练的结果。观察发现，当采用较大的固定功率训练时，EMNN更注重高发射功率下的表现效果（如图中15dBm-30dBm），而在低发射功率条件下表现不佳。对应的，当采用较小的固定功率训练时，EMNN更注重低发射功率下的表现效果（如图中0dBm-10dBm）,而在高发射功率条件下表现不佳。然而，当EMNN使用Beta分布随机生成的功率训练时，可以同时兼顾高发射功率和低发射功率的表现效果，从而整体上实现更低的BER。此外，结合Fig.~\ref{Fig9}，我们发现基于迁移学习的训练方法能够实现与直接训练相当的效果，同时显著加快训练速度。
As shown in Fig.~\ref{Fig8}, we compare the performance of the EMNN trained with different strategies. It can be observed that when trained with a larger fixed transmit power, the EMNN tends to focus on performance at high transmit powers (e.g., 15~dBm--30~dBm), but performs poorly under low transmit power conditions. Conversely, when trained with a smaller fixed transmit power, the EMNN emphasizes low-power performance (e.g., 0~dBm--10~dBm), but degrades at higher transmit powers. In contrast, when the EMNN is trained with transmit powers randomly generated according to a Beta distribution, it can simultaneously maintain robust performance under both high and low transmit powers, thereby achieving an overall lower BER. Furthermore, as shown in Fig.~\ref{Fig9}, we observe that the transfer learning-based training method can achieve comparable performance to direct training, while significantly reducing the training time.

\begin{figure}[!t]
\centering
\includegraphics[width=2.85in]{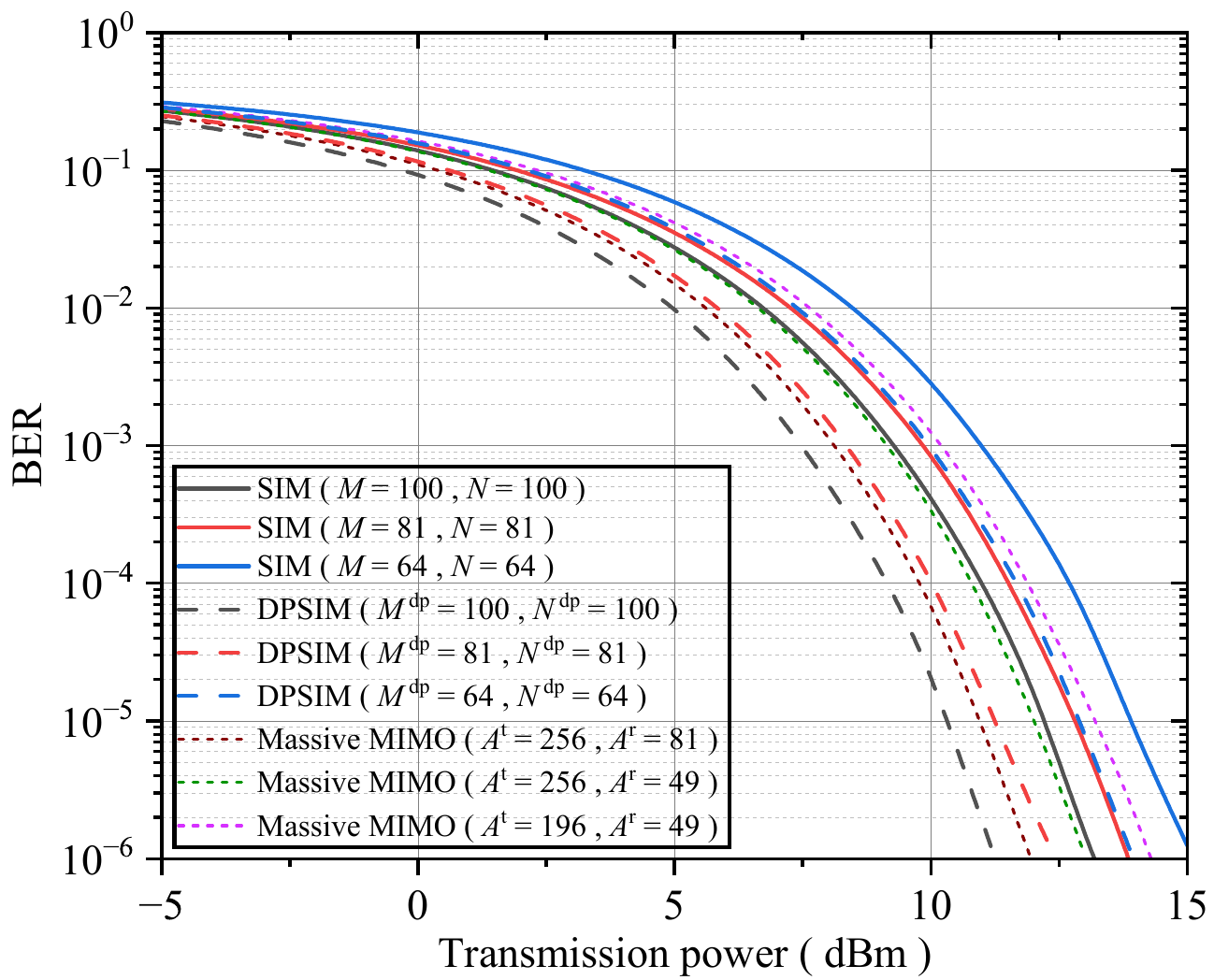}
\caption{Performance comparison of different EM (DPEM) unit quantities.}
\label{Fig10}
\vspace{-0.1cm}
\end{figure}

\begin{figure}[!t]
\centering
\includegraphics[width=2.85in]{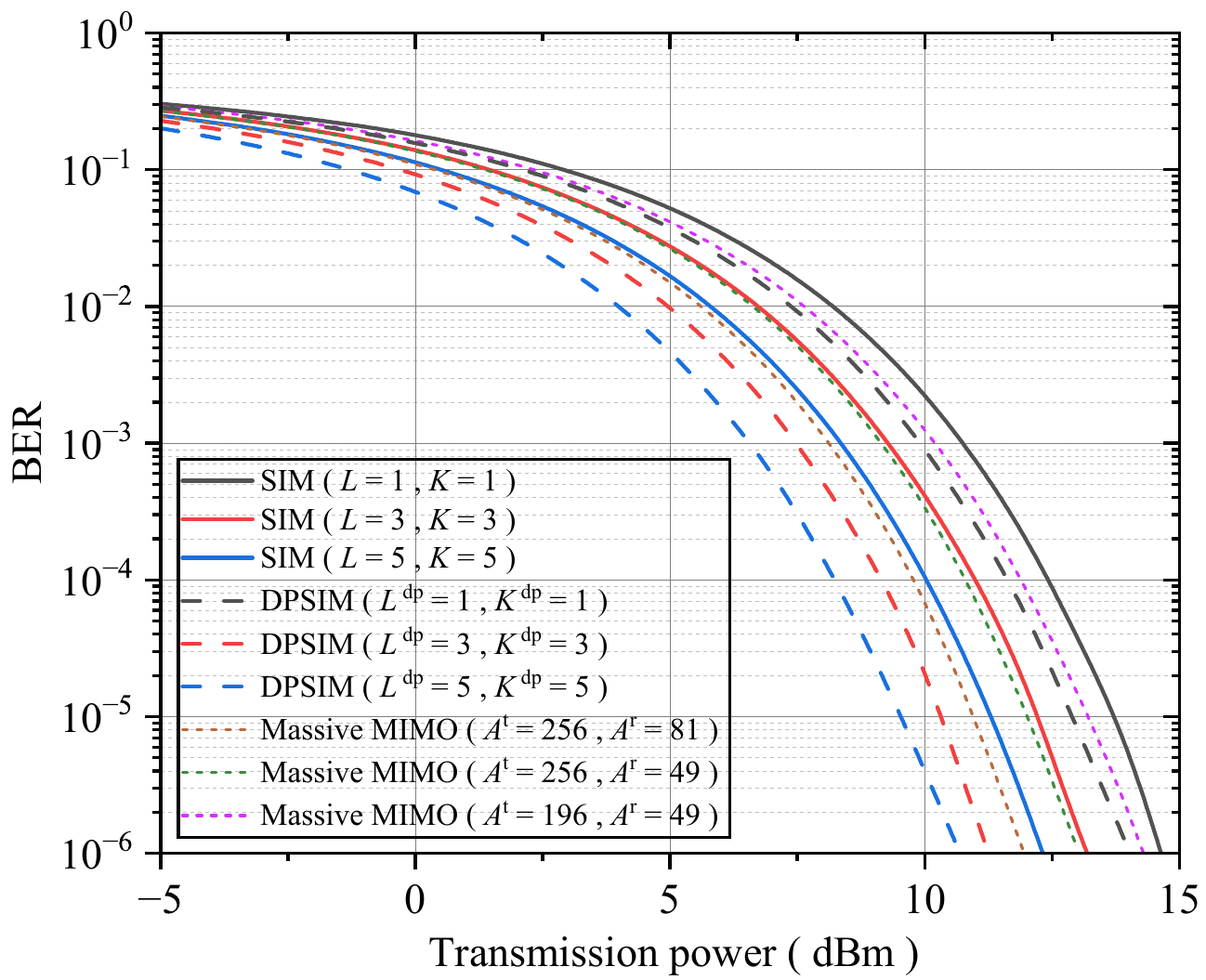}
\caption{Performance comparison of different numbers of metasurface layers.}
\label{Fig11}
\end{figure}

\subsection{Performance comparison analysis of different specifications of SIM (DPSIM).} \label{Ⅳ.~C}
\begin{figure}[!t]
\centering
\includegraphics[width=2.85in]{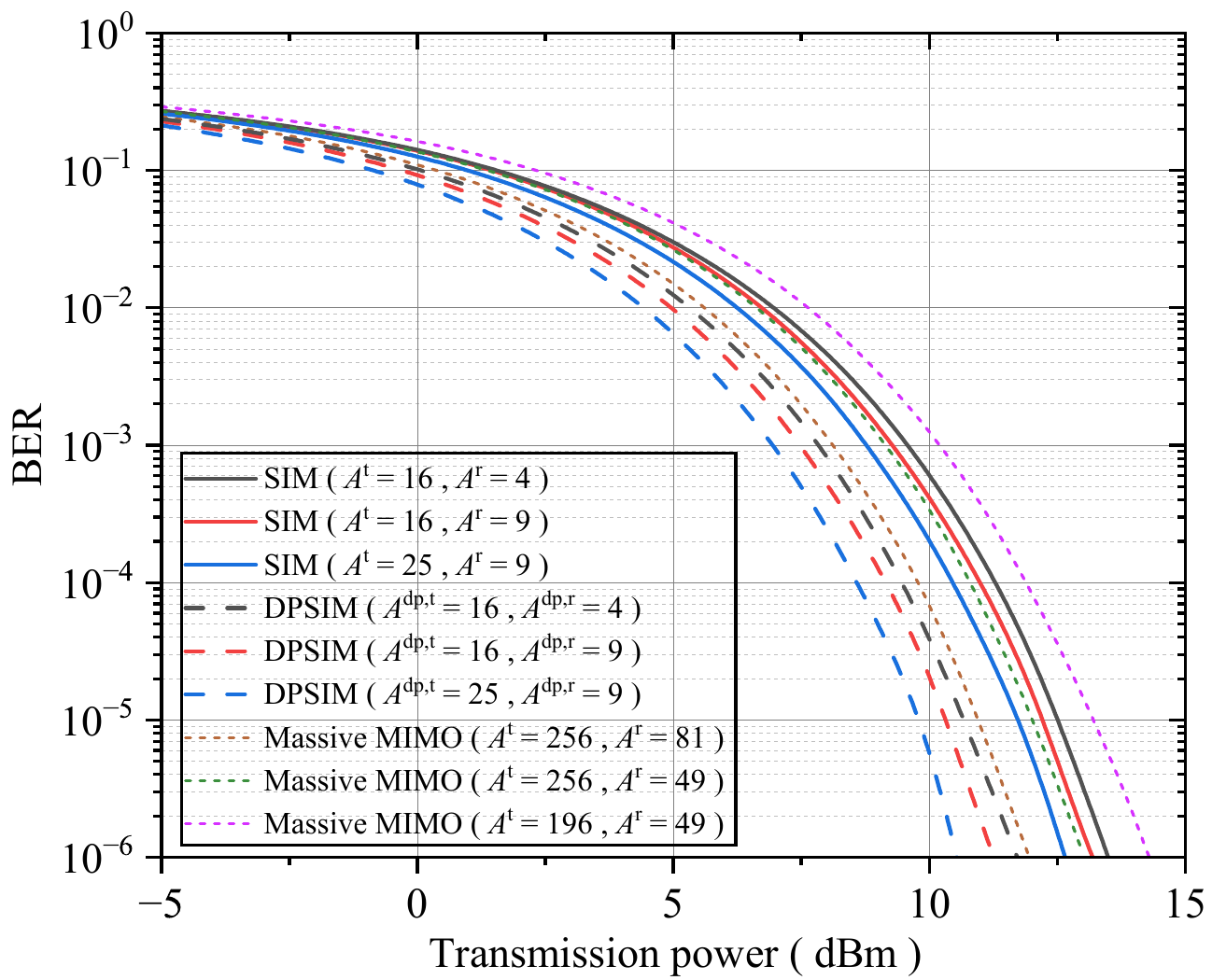}
\caption{Performance comparison of different antenna quantities.}
\label{Fig12}
\vspace{-0cm}
\end{figure}

Fig.~\ref{Fig10} -- Fig.~\ref{Fig12} depict the BER curves with varying numbers of EM (DPEM) units, metasurface layers, and transceiver antennas, respectively. It can be observed that enlarging the EM (DPEM) array size, increasing the number of metasurface layers, and employing more transmit (receive) antennas all contribute to effectively reducing the system BER. Moreover, under all configurations, the DPSIM-assisted E2E OFDM system consistently achieves significantly better BER performance than its SIM-assisted counterpart. Particularly, we observe that by integrating the SIM (DPSIM) device, an E2E OFDM system with 16T9R can achieve comparable performance to a massive MIMO OFDM system with 256T49R. This validates the potential of incorporating SIM (DPSIM) devices for wave-domain signal processing to offload baseband processing tasks, thereby simplifying the transceiver design and reducing hardware cost.
%Fig.\ref{Fig10}-Fig.\ref{Fig12}依次展示了不同EM (DPEM) 单元数量、超表面层数以及收发天线数量的 BER 曲线。可以观察到，扩大EM(DPEM)单元规模、增加超表面层数、提高收发天线数量，均能够有效降低系统的BER。此外，在各类配置下，DPSIM 辅助E2E OFDM系统的 BER 性能均显著优于 SIM 辅助E2E OFDM系统。 特别的，我们发现通过集成SIM （DPSIM）器件，16T9R的E2E OFDM系统可以与256T49R的Massive MIMO OFDM系统性能相当。这验证了通过集成SIM (DPSIM)器件在波域进行信号处理以卸载基带处理任务，最终简化收发机和降低硬件成本的可能性。

\begin{figure}[!t]
\centering
\includegraphics[width=2.85in]{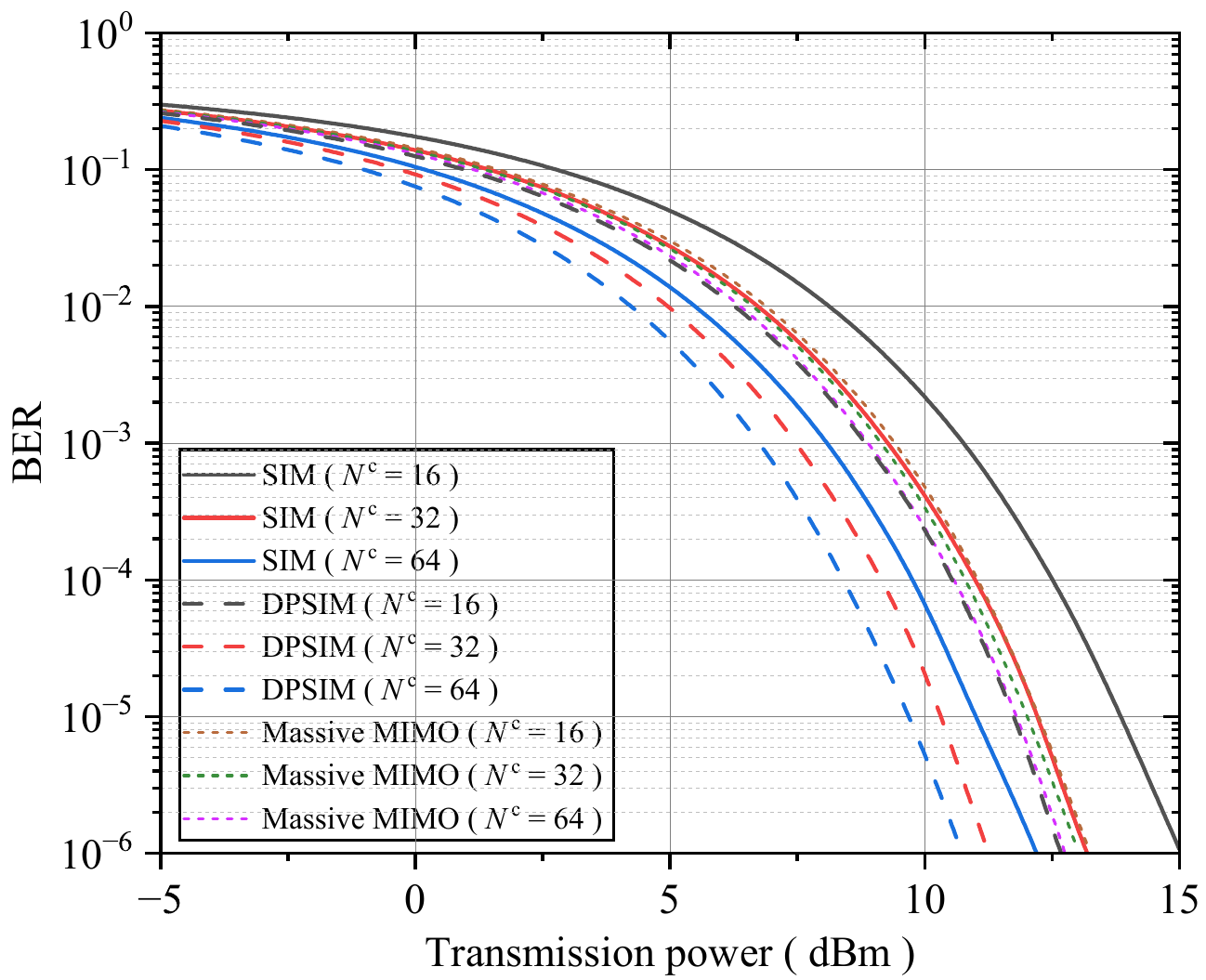}
\caption{Performance comparison of different subcarrier quantities. Massive MIMO (256T49R).}
\label{Fig13}
\end{figure}

\subsection{Comparative analysis of different numbers of subcarriers, transmission bit lengths and polarization defect intensities.} \label{Ⅳ.~D}
\begin{figure}[!t]
\centering
\includegraphics[width=2.85in]{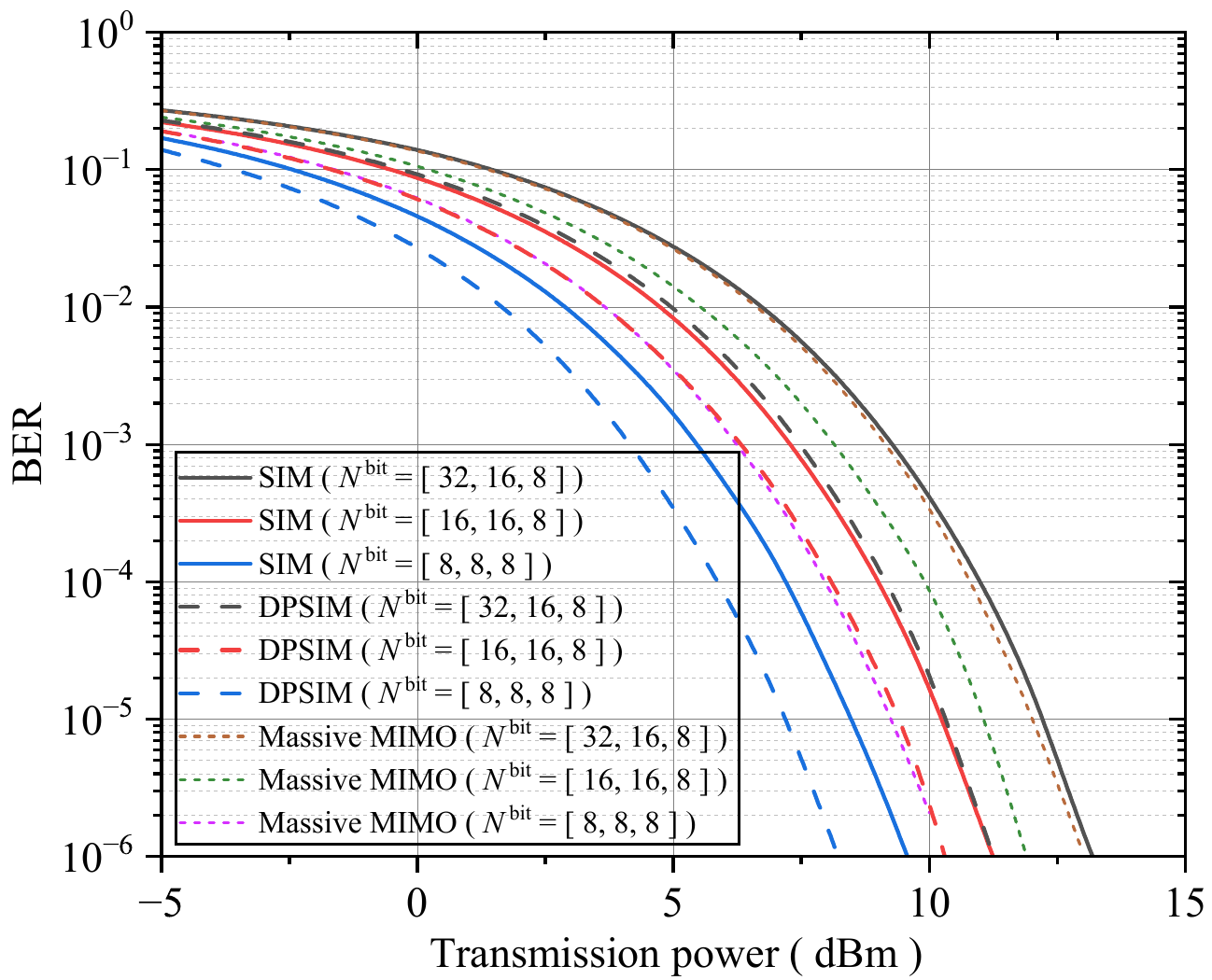}
\caption{Performance comparison of different transmission bit numbers. Massive MIMO (256T49R).}
\label{Fig14}
\vspace{-0.1cm}
\end{figure}

\begin{figure}[!t]
\centering
\includegraphics[width=2.85in]{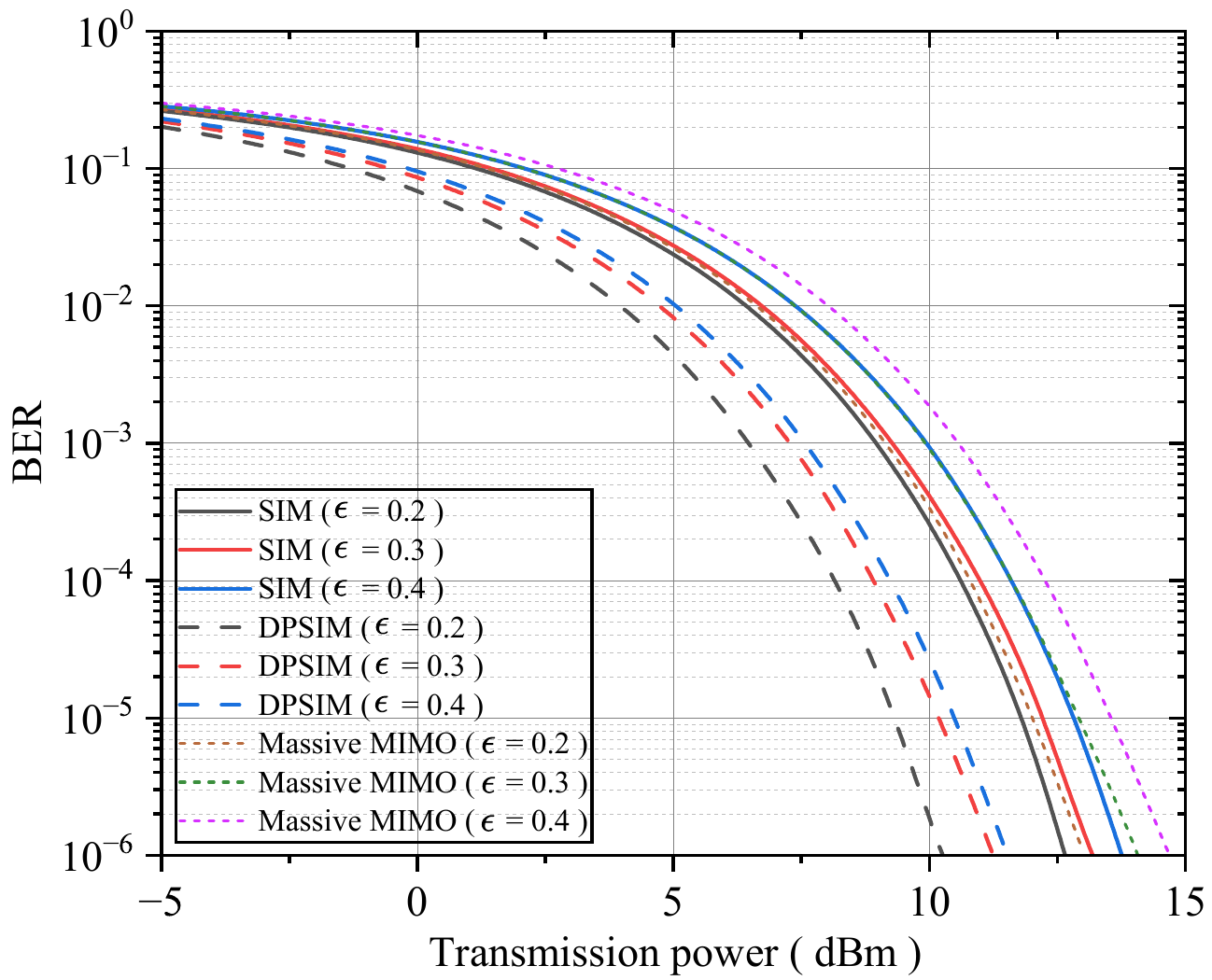}
\caption{Performance comparison of different polarization radiation factors. Massive MIMO (256T49R).}
\label{Fig15}
\end{figure}

%Fig.\ref{Fig13}给出了不同子载波数量下的 BER 曲线。结果表明，子载波数量增加可以有效降低系统BER，且 DPSIM 辅助E2E OFDM系统在性能上显著优于 SIM 辅助系统。Fig.\ref{Fig14}展示了在一个 OFDM 符号周期内，三个用户各自采用不同比特位数传输时的 BER 曲线。可以看到，当一个 OFDM 符号对应的比特总数减少时，系统BER随之降低，且在各情形下，DPSIM 辅助E2E OFDM系统均优于 SIM 辅助系统。 Fig.\ref{Fig15}给出了不同极化交叉因子条件下的 BER 曲线。结果显示，当极化交叉因子 $\epsilon$ 越小（即不同极化方向之间的能量转换越少）时，系统BER越低。由于 DPSIM 具备同时处理两个极化方向信号的能力，其辅助E2E OFDM系统在此类场景下相较 SIM 辅助E2E OFDM系统表现出更明显的性能优势。 
Fig.~\ref{Fig13} presents the BER performance with different numbers of subcarriers. The results indicate that increasing the number of subcarriers can effectively reduce the system BER, and that the DPSIM-assisted E2E OFDM system achieves a significant performance improvement compared to the SIM-assisted counterpart. Fig.~\ref{Fig14} shows the BER curves for three users within one OFDM symbol period when each user transmits a different number of bits. It can be observed that as the total number of bits per OFDM symbol decreases, the system BER decreases accordingly. In all the considered cases, the DPSIM-assisted E2E OFDM system consistently outperforms the SIM-assisted system. Fig.~\ref{Fig15} illustrates the BER performance under different radiated power conversed factors. The results demonstrate that a smaller $\epsilon$ (i.e., less energy coupling between orthogonal polarization directions) leads to a lower system BER. Owing to its capability of simultaneously processing signals from two polarization directions, the DPSIM-assisted E2E OFDM system exhibits more pronounced performance gains over the SIM-assisted counterpart in such scenarios.

\section{Conclusion.} \label{Ⅴ}
%为进一步提高SIM (DPSIM)辅助通信系统的整体性能，我们提出了SIM (DPSIM)辅助的E2E OFDM系统架构，并设计了一种基于迁移学习的EMNN训练和部署框架，用于发送比特流到接收比特流全链路联合优化。仿真结果表明，在复杂信道条件下，SIM辅助和DPSIM辅助的E2E OFDM系统均能实现稳健的比特流传输。值得注意的是，DPSIM辅助系统表现出相较SIM辅助系统更优的误码率（BER）性能。通过将波域信号处理与基于深度学习的控制相结合，所提出的系统显著减少了对数字基带处理能力的依赖，从而简化了收发机设计并提高了能量效率。本研究凸显了EMNN与SIM (DPSIM)辅助E2E OFDM系统在下一代智能收发机设计中的巨大潜力。
To further enhance the overall performance of SIM (DPSIM)- assisted communication systems, we propose an SIM (DPSIM)- assisted E2E OFDM system architecture and design a transfer-learning-based EMNN training and deployment framework for E2E joint optimization from transmitted bit streams to received bit streams. Simulation results demonstrate that, under complex channel conditions, both the SIM-assisted and DPSIM-assisted E2E OFDM systems can achieve robust bit-stream transmission. It is noteworthy that the DPSIM-assisted system exhibits superior BER performance compared with its SIM-assisted counterpart. By integrating wave-domain signal processing with deep-learning-based control, the proposed system significantly reduces the reliance on digital baseband processing capability, thereby simplifying transceiver design and improving energy efficiency. This study highlights the great potential of EMNN combined with SIM (DPSIM)-assisted E2E OFDM systems for next-generation intelligent transceiver design.

\bibliographystyle{IEEEtran}
\bibliography{references}

\end{document}